\documentclass{article}

\usepackage{arxiv}

\usepackage{amsmath}
\usepackage[utf8]{inputenc} 
\usepackage[T1]{fontenc}    
\usepackage{hyperref}       
\usepackage{url}            
\usepackage{booktabs}       
\usepackage{amsfonts}       
\usepackage{nicefrac}       
\usepackage{microtype}      
\usepackage{graphicx}
\usepackage[numbers]{natbib}
\usepackage{doi}

\usepackage[english]{babel}
\usepackage{xcolor}
\usepackage[ruled,vlined]{algorithm2e}

\usepackage{caption}
\usepackage{subcaption}

\usepackage{ifthen}
\usepackage{url}

\usepackage{booktabs}
\usepackage{hyphenat}
\usepackage{multirow}

\usepackage{hypernat}

\usepackage{xspace}
\usepackage{wrapfig}
\usepackage{pifont}
\usepackage[nohyperlinks]{acronym}
\usepackage[abbreviations]{foreign}  

\graphicspath{ {./images/} }

\usepackage[toc,page]{appendix}

\DeclareUnicodeCharacter{202F}{}

\newcommand{\sysname}{\textsc{PEPPER}\xspace}

\LinesNumbered 
\DontPrintSemicolon 
\usepackage{algcompatible}
\renewcommand{\algorithmiccomment}[1]{\bgroup\hfill$\rhd$\small~#1\egroup}

\SetFuncSty{sc}
\SetDataSty{em}
\SetCommentSty{em}
\SetArgSty{}

\SetKwFor{func}{function}{}{end}
\SetKwFor{upon}{upon}{}{end}
\SetKwFor{struct}{struct}{}{end}

\SetKw{Break}{break}

\SetKwBlock{Block}{}{}

\usepackage{tikz}
\usepackage{xcolor}
\newcommand{\cercle}[1]{\tikz[baseline=(myanchor.base)] \node[circle,fill=.,inner sep=1pt] (myanchor) {\color{-.}\bfseries\footnotesize #1};}

\title{\sysname: Empowering User-Centric Recommender Systems over Gossip Learning}

\author{ \href{https://orcid.org/0000-0003-3433-1337}{\includegraphics[scale=0.06]{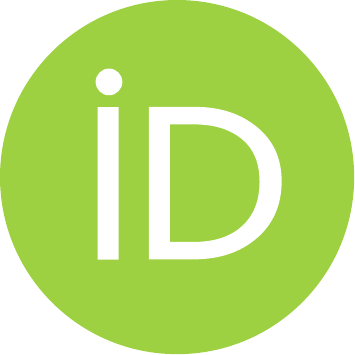}\hspace{1mm}Yacine Belal}\\
	INSA Lyon \\
	LIRIS \\
	Lyon, France \\
	\texttt{yacine.belal@insa-lyon.fr} \\
	\And
	\href{https://orcid.org/0000-0003-3440-1251}{\includegraphics[scale=0.06]{orcid.pdf}\hspace{1mm}Aurélien Bellet} \\
	Univ. Lille \\
	INRIA \\
	CNRS\\
	Lille, France \\
	\texttt{aurelien.bellet@inria.fr} \\
	\And 
	\href{https://orcid.org/0000-0003-2821-7714}{\includegraphics[scale=0.06]{orcid.pdf}\hspace{1mm}Sonia Ben Mokhtar} \\
	INSA Lyon \\
	LIRIS \\
	CNRS \\
	Lyon, France \\
	\texttt{sonia.benmokhtar@insa-lyon.fr} \\
	\And 
	\href{https://orcid.org/0000-0002-7996-3963}{\includegraphics[scale=0.06]{orcid.pdf}\hspace{1mm}Vlad Nitu} \\
	INSA Lyon \\
	LIRIS \\
	CNRS \\
	Lyon, France \\
	\texttt{vlad.nitu@cnrs.fr} \\
}

\begin{document}
\maketitle

\begin{abstract}
Recommender systems are proving to be an invaluable tool for extracting user-relevant content helping users in their daily activities (e.g., finding relevant places to visit, content to consume, items to purchase). However, to be effective, these systems need to collect and analyze large volumes of personal data (e.g., location check-ins, movie ratings, click rates .. etc.), which exposes users to numerous privacy threats. In this context, recommender systems based on Federated Learning (FL) appear to be a promising solution for enforcing privacy as they compute accurate recommendations while keeping personal data on the users' devices. However, FL, and therefore FL-based recommender systems, rely on a central server that can experience scalability issues besides being vulnerable to attacks. To remedy this, we propose \sysname, a decentralized recommender system based on gossip learning principles. In \sysname, users gossip model updates and aggregate them asynchronously. At the heart of \sysname reside two key components: a personalized peer\hyp{sampling} protocol that keeps in the neighborhood of each node, a proportion of nodes that have similar interests to the former and a simple yet effective model aggregation function that builds a model that is better suited to each user. Through experiments on three real datasets implementing two use cases: a location check-in recommendation and a movie recommendation, we demonstrate that our solution converges up to 42\% faster than with other decentralized solutions providing up to 9\% improvement on average performance metric such as hit ratio and up to 21\% improvement on long tail performance compared to decentralized competitors.

\end{abstract}


\section{Introduction}

Recommender systems are at the heart of many popular online services used by billions of users on a daily basis to get insights on what could be a good restaurant to eat in, a nice museum to visit or the next TV series to watch. For instance, around 53 million US users perform smartphone searches on a daily basis about local businesses/services in their immediate surroundings and visit these places~\cite{Statista}. 
To be effective, recommender systems generally require the collection of large corpuses of personal data (e.g., check-ins, clicks, ratings) and substantial computing power in order to be trained.
Therefore, they are traditionally centralized and often hosted in the data center of the service provider. 
However, this approach poses serious privacy concerns due to the never-ending list of attacks and privacy scandals~\cite{forbes2020,cbsnews2019,ciscomag2019}, which continue to unfold.
These scandals generally harm the image of the service provider and cause strong reactions in the public opinion.
Furthermore, surveyed users generally consider that (centralized) recommender systems violate their privacy~\cite{mohallick2018towards} and would prefer not to be profiled~\cite{awad2006personalization}.

In the past decade, there have been various solutions investigating privacy-preserving recommender systems~\cite{gao2020,wang2019,shin2018,guerraoui2017,Privacy-preservingdistributedcollaborativefiltering, mcsherry2009}. 
Among these solutions, Federated Recommender Systems, i.e., recommender systems leveraging the Federated Learning principles (FL), are considered promising solutions ~\cite{Guo21,muhammad2020fedfast,ASimpleandEfficientFederatedRecommenderSystem, Fast-adaptingandPrivacy-preservingFederatedRecommender} as they provide privacy-by-design guarantees. Indeed, instead of centralizing users' personal data and training recommendation models on remote cloud infrastructures, FL proposes to train the models right where the data is generated, i.e., on the edge / mobile devices. However, this solution has its own drawbacks due to the inherent limitations of FL, one such limitation being its centralized master-slave architecture.
Even if FL allows users to keep their data local, they still send ML model updates to a (logically) central server, which drives the model convergence process.
This centralized server poses not only fault tolerance but also scalability problems since it restricts Federated Recommender Systems only to those companies which are able to scale the centralized server up to the necessary number of clients~\cite{CanDecentralizedAlgorithmsOutperformCentralizedAlgorithms?ACaseStudyforDecentralizedParallelStochasticGradientDescent}. To address this issue, we investigate in this paper decentralized recommender systems over Gossip Learning~\cite{DecentralizedRecommendationBasedonMatrixFactorization:AComparisonofGossipandFederatedLearning,Decentralizedlearningworks:Anempiricalcomparisonofgossiplearningandfederatedlearning}.
In Gossip Learning, nodes exchange model updates with their neighbors and asynchronously aggregate the received models using a model aggregation function (e.g., federated averaging~\cite{FederatedLearningofDeepNetworksusingModelAveraging}).
There exist preliminary solutions for decentralized recommender systems over Gossip Learning~\cite{DecentralizedRecommendationBasedonMatrixFactorization:AComparisonofGossipandFederatedLearning}. However, these solutions suffer from poor average, and even worse tail performance compared to their centralized counterpart. Said differently, existing solutions tend to improve the average user satisfaction at the cost of having extremely unsatisfied users at the long and short tail.

To address this issue, we present \sysname, a novel solution that aims at maximizing average user performance without penalizing users at the tail. \sysname is composed of a performance-based model aggregation function and a personalized peer-sampling protocol. While the former compares the received models with respect to their performance on a local test set and gives more weight to those models that perform better from the point of view of each user, the latter keeps a proportion of seemingly similar neighbors in the view of each user, \ie, neighbors that previously sent models, which performed well for this user. We evaluate \sysname with extensive simulations involving up to one thousand nodes and by relying on two different machine learning models trained on three real world datasets: Foursquare\hyp{NYC}, Gowalla\hyp{NYC}~\cite{snapnets} and MovieLens ML-100k~\cite{movielens} and compare its performance against six state-of-the-art federated and decentralized solutions. Our results show that \sysname systematically improves the average performance of decentralized recommender systems (up to 9$\%$, 6$\%$ and 13$\%$ improvement on performance metrics such as Hit Ratio, NDCG, F1 score, respectively) and in some cases outperforms the average performance of its centralized counterparts.
In addition to improving average performance, \sysname substantially improves long tail performance compared to both federated and decentralized competitors. Finally, thanks to its personalized peer-sampling protocol, \sysname converges up to 42$\%$ faster than other decentralized competitors. \\

The rest of the paper is organised as follows. In Section~\ref{sec:background}, we present a background on Federated Learning, Gossip Learning and the recommendation models we rely on in this work. In Sections~\ref{sec:overview} and \ref{sec:details}, we present an overview of \sysname and its detailed description. We then present the performance evaluation of \sysname in Section~\ref{sec:eval} and discuss its limitations and possible mitigations in Section~\ref{sec:discussion}. Finally, we present the related research works in Section~\ref{sec:rel_works} and conclude the paper in Section~\ref{sec:conclusion}.

\section{Background}
\label{sec:background}

We present in this section preliminary background on key concepts related to \sysname, namely Federated Learning (Section~\ref{subsec:fl}), Gossip Learning (Section~\ref{subsec:gl}) and the used recommendation models (Section~\ref{subsec:recsys}).

\subsection{Federated Learning}
\label{subsec:fl}
\begin{figure}
\centering
\begin{subfigure}[b]{0.5\textwidth}
    \centering
    \includegraphics[width=\textwidth]{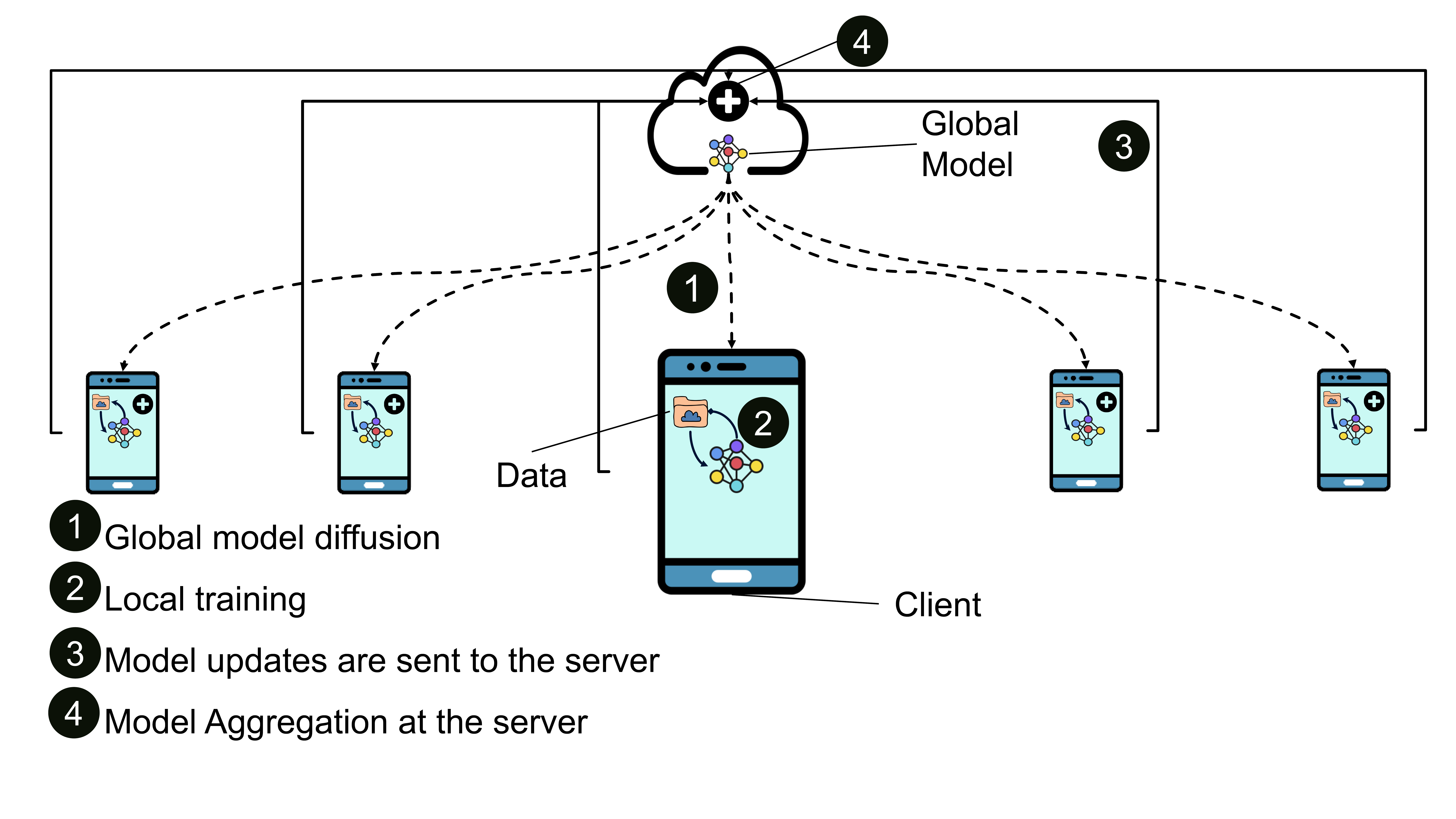}
    \caption{Federated Learning Architecture~\cite{FederatedLearningofDeepNetworksusingModelAveraging}}
    \label{fig:fl}
\end{subfigure}
\hfill
\begin{subfigure}[b]{0.45\textwidth}
    \centering
    \includegraphics[width=\textwidth]{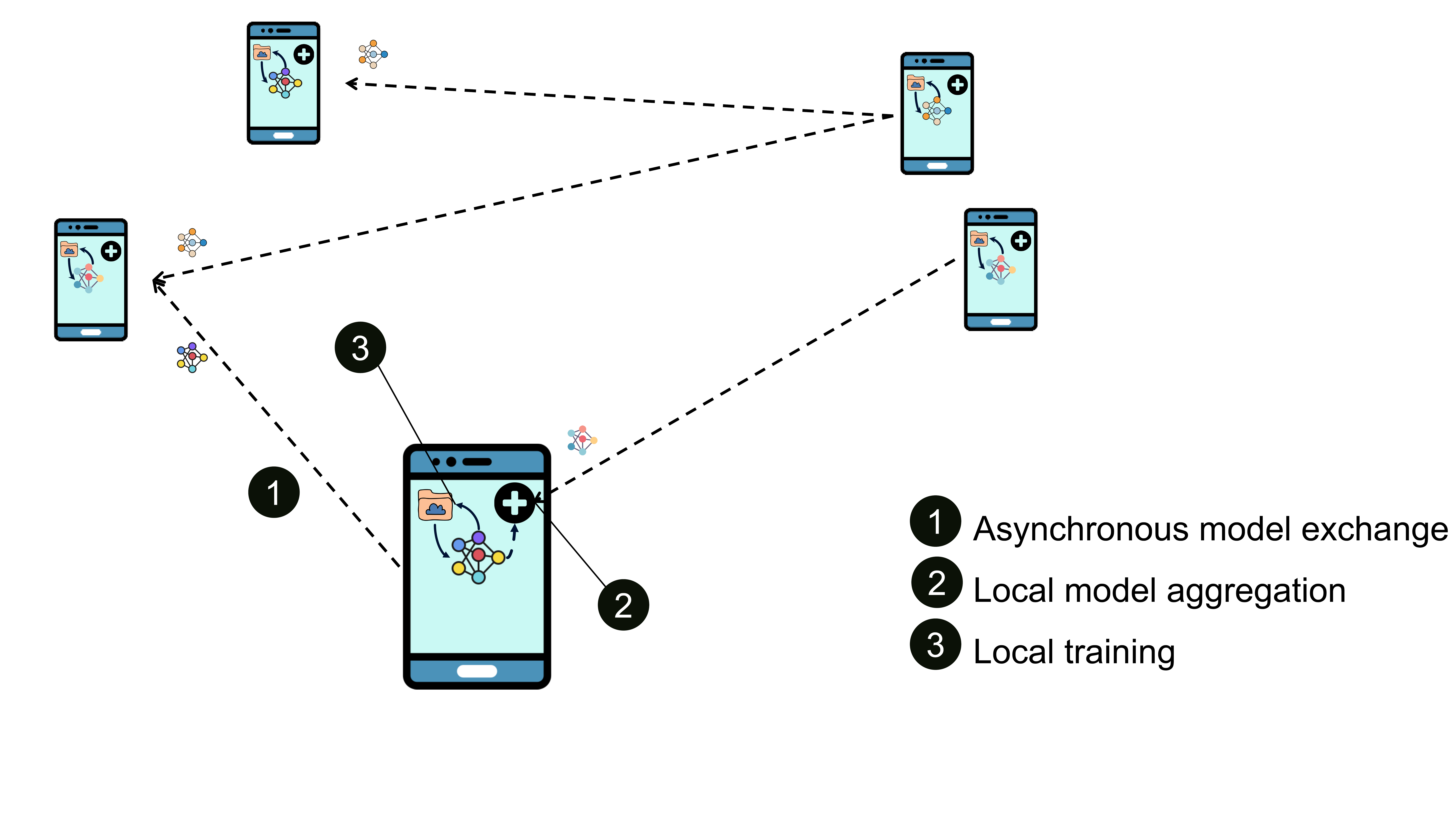}
    \caption{Gossip Learning Architecture}
    \label{fig:gl}
\end{subfigure}   
\caption{Federated Learning vs Gossip Learning Architectures.}
\end{figure}

Federated Learning (FL), depicted in Figure~\ref{fig:fl}, is a computing paradigm that instantiates the "bring code close to the data" principle in the context of Machine Learning.
It was first proposed by Google~\cite{FederatedLearningofDeepNetworksusingModelAveraging} to allow hundreds of millions of mobile devices (clients) to collaboratively train and build a global model while preserving the privacy of their data.  
To this end, FL assumes the presence of a coordinator (i.e, a logically centralized server) and $N$ clients $C_i$, where $i \in [1,N]$. Each client $C_i$ creates and stores data locally, say $D_i$. Given a learning task, let us denote by $M_{global}$ the global model held by the coordinator and by $M_{i}$ the model held by each client $C_i$. Following an initialization phase where the coordinator communicates the model architecture and the optimization algorithm to clients, the former continuously initiates training rounds until the convergence of the global model is considered as satisfactory. Each round is conducted as follows: (i) the coordinator randomly selects a fraction $p$ of clients of the original client set and sends them the current state of the global model $M_{global}$ (see step \cercle{1} in Figure~\ref{fig:fl}); (ii) each client $C_i$ initializes its model $M_i$ with the model parameters received from the coordinator. $M_i$ is then trained on the local dataset $D_i$ before being sent back to the coordinator (see steps \cercle{2} and \cercle{3} in Figure~\ref{fig:fl}); (iii) the coordinator aggregates the different models' parameters received and updates the global model (see step \cercle{4} in Figure~\ref{fig:fl}). A classical way of aggregating models in FL is by using the \emph{FedAvg}~\cite{FederatedLearningofDeepNetworksusingModelAveraging} algorithm, which relies on Formula~\ref{eq:FedAvg}. \emph{FedAvg} computes a weighted average of the received models, parameter wise. Specifically, each model is weighted by the number of samples it was trained on, that is, the dataset size of the owner of the model (\cf, equation \eqref{eq:FedAvg}).

\begin{equation}\label{eq:FedAvg}
    M_{global} = \frac{1}{\sum_{i=1}^{p} \lvert{D_i}\rvert} \sum_{i=1}^{p} \lvert{D_i}\rvert {M_i} 
\end{equation}

While it enhances data privacy, FL is known to have convergence properties that slightly deviate from classical centralized ML~\cite{FederatedMachineLearning:ConceptandApplications}. Additionally, the FL architecture heavily relies on the presence of a central server, which plays a key role in the orchestration of the learning process, both at the start of each round and during the aggregation phase. The necessity of this server has been often criticized as it makes FL vulnerable to failures, scalability issues~\cite{AdvancesandOpenProblemsinFederatedLearning} and attacks~\cite{CanYouReallyBackdoorFederatedLearning?,AttackoftheTails:YesYouReallyCanBackdoorFederatedLearning}.
This has quite naturally created within research works a tendency towards approaches that relax the single server assumption. In the next section, we present a gossip-based approach that we adopted in this work.

\subsection{Gossip Learning}
\label{subsec:gl}
Gossip Learning is an asynchronous learning protocol that was first proposed by~\citet{ormandi2011} to solve the problem of learning over fully distributed data using a peer-to-peer communication protocol called gossip. Its flexibility, privacy-preserving nature and scalability properties give it the potential to be a first-class citizen in the field of distributed machine learning. 
Gossip Learning relies on a key protocol called \emph{peer\hyp{sampling}} that provides every node with a set of peers, called a view, to gossip with. The view of each node is periodically shuffled so that nodes have the opportunity to interact with new nodes in the system. This protocol plays a crucial role in the dissemination speed of messages. In gossip learning, it can even impact the performance of the learned models as further illustrated in Section~\ref{sec:eval}.
Indeed, the peer\hyp{sampling} protocol enables each node to periodically change partially or totally its view. This change can be done randomly, creating an unstructured network. It can also follow a well defined logic that can speed-up the convergence time, reduce the communication overhead or improve the performance of the overall system~\cite{Apeer-to-peerrecommendersystemforself-emergingusercommunitiesbasedongossipoverlays}. 

In a classical Gossip Learning algorithm, a node periodically sends its local model to its view $P$. Then, upon the reception of a model $m$ from one of its neighbors, it aggregates this model according to a predefined \emph{model aggregation function}. Later on, the obtained model is updated by performing a local training using a local dataset $D$. The resulting model is thereafter considered as the new current model. In the above algorithm, the model aggregation function is a key component that makes it possible for nodes to learn from others nodes' data without having to actually train on it. This makes it one of the most crucial components in Gossip Learning. Model aggregation functions generally perform a weighted averaging of the models' parameters or a subset of these parameters. Algorithm~\ref{alg:aggregationmethods} illustrates two common examples of model aggregation functions, namely Decentralized FedAvg~\cite{fullavg} and Model-Aged-Based~\cite{Decentralizedlearningworks:Anempiricalcomparisonofgossiplearningandfederatedlearning}. While the former applies the FedAvg algorithm defined in FL to pairs of models, the latter is more advanced as it is based on a notion of model age, which reflects how much a model has circulated in the network. The intuition behind it is to give more weight to older models as they are likely to hold more knowledge.

\begin{algorithm}[!h]
\DontPrintSemicolon

\SetKwFunction{fedavg}{Decentralized FedAvg}
\SetKwFunction{modelagebased}{ModelAgeBased}
\SetKwProg{Fn}{Procedure}{:}{}

\KwData{train data sizes $D_1$, $D_2$; models to be aggregated : $m_1$,$m_2$; models' number of updated times : $age_1$, $age_2$;}
\Fn{\fedavg{$m_1$, $m_2$}}{
 
 \KwRet $ \frac{1}{\lvert{D_1}\rvert + \lvert{D_2}\rvert} (\lvert{D_1}\rvert \times {m_1}) + (\lvert{D_2}\rvert \times {m_2})  $
  }
  \;
 
\Fn{\modelagebased{$m_1$, $m_2$}}{
 \KwRet $ \frac{1}{\lvert{age_1}\rvert + \lvert{age_2}\rvert} (\lvert{age_1}\rvert \times {m_1}) + (\lvert{age_2}\rvert \times {m_2})  $
  }
  \;

\caption{Aggregation Function : Implementations}
\label{alg:aggregationmethods}
\end{algorithm}

\subsection{The Recommendation Models}
\label{subsec:recsys}
While \sysname is agnostic to the underlying machine learning model, we use two models in our experiments: the Generalized Matrix Factorization (GMF) model proposed by~\citet{NCF} and the Personalized Ranking Metric Embedding Method (PRME-G) proposed by~\citet{feng2015personalized}. This choice is motivated by two criteria: (i) the effectiveness of these models, which have a sufficiently accurate performance to allow us to quantify the contribution of our system and compare it with different competitors and (ii) their lightness. Indeed, compared to other benchmark recommendation models, GMF and PRME-G are light enough to be considered in a  Gossip Learning context where nodes can have constrained resources.
Concretely, GMF~\cite{NCF} is a neural network inspired by matrix factorisation (MF). As input, this model takes both user and item identifiers. It is followed by an embedding layer that projects them into a latent vector space. These vectors are equivalent to the latent feature vectors of MF. A multiplication layer does a point-wise product on these two vectors, to combine the features describing the item and those describing the user. The product is then fed to at least one fully connected layer before passing through a sigmoid activation function that outputs the degree of relevance of the item to the user. By comparing this output with the actual relevance of the item, GMF computes an error function (\ie, binary cross entropy) and optimizes the embedding features, until it finds features that best describe the user's preferences and the item's characteristics.

As opposed to GMF, PRME-G~\cite{feng2015personalized} is specific to the point\hyp{of}\hyp{interest} recommendation task. By using the metric embedding method, it puts locations in a latent space and minimizes the Euclidean distance between each pair of locations $(x,y)$ the more likely is $y$ to be visited after $x$. Similarly, it minimizes in another latent space the distance between users and locations, based on their preferences. Finally, it incorporates spatial influence by giving more weight to closer POIs. These three pieces of information are combined into a three dimensional matrix $D$ where each cell $(u,l^s,l_i)$ is the likelihood of user $u$ visiting location $l_i$ starting from location $l^s$. To optimize this matrix, for each training observation $(u,l^s,l_i)$ a random POI $l_j$ that is not observed for $(u,l^s)$ is generated and the following update rule is executed:
\begin{equation}
    \Theta \leftarrow \Theta + \gamma  \frac{\partial}{\partial \Theta} (\log(\sigma(z))) - \lambda \lVert \Theta \rVert ^2 ) 
\end{equation}

\noindent where $z = D_(u,l^s,l_j)$ - $D_(u,l^s,l_i)$ and $\gamma$, $\sigma$ and $\Theta$ are respectively the learning rate, the sigmoid function, and the matrices representing locations' and/or users' latent representations.

\section{\sysname in a Nutshell}
\label{sec:overview}
In this section, we start by defining the objectives of \sysname (Section~\ref{subsec:objectives}) and our assumptions (Section~\ref{subsec:assumptions}) before presenting an overview of our system (Section~\ref{subsec:overview}).

\subsection{\sysname Objectives}
\label{subsec:objectives}
Existing decentralized solutions have a centralized perspective of the learning process. Indeed, they aim at building $N$ models that best approximate the global data distribution, that is, $N$ models converging to a similar minimum as a global FL model. 
They do so by aggregating models generically and evaluating them on global metrics, which quantify the ability of models to perform well for any given user. 
While building such models seems to be the natural way to go for decentralizing FL, it may lead to poor performance because of the following reasons: (i) users can have different distributions which can lead to a phenomenon called client drift~\cite{AdvancesandOpenProblemsinFederatedLearning} that can hurt convergence and (ii) targeting the same minima in a decentralized system is costly and requires aggregating a large number of models, hence, a lot of communication messages. 
In this work, we argue that these limitations can be circumvented in use cases where approximating a global distribution is not necessary to achieve user satisfaction. 
For instance, in a recommendation system, users are mainly interested in having a model that recommends items which are  highly relevant to them and may not care much about the model's ability to recommend items for other users. 
This can be achieved by approximating a local yet more relevant distribution for each model. 
In this context, the main objective of \sysname is to train $N$ decentralized recommendation models that are each tailored to their respective user. 
More specifically, we target a better average personalized-data performance, which is described in~\cite{Schneider2019} as the ability of a model to derive the same decisions as specified by the individual, even if these decisions do not hold with respect to the performance metrics evaluated on a global dataset. 
As a positive side effect, this can also improve the performance of tail users, that are often  neglected in a global metric optimization approach.  

\subsection{\sysname Assumptions}
\label{subsec:assumptions}
In \sysname, we make several assumptions, some of them being common in decentralized ML systems. 
First, we are only concerned with the performance aspects of gossip-based recommender systems. Therefore, we do not quantify its robustness to attacks nor do we consider the presence of any malicious or curious users. All users are assumed to be honest clients aiming to train a model that has the best possible local performance.
We assume that nodes have heterogeneous bandwidth connectivity and may experience networking delays. However, we assume reliable communication links (i.e., if a message is lost it is retransmitted sufficiently enough to be received) and a stable participation of nodes in the system (i.e., no churn).
Then, as commonly considered in decentralized ML systems, we also assume the presence of a central bootstrap service in charge of broadcasting to all nodes the optimization algorithm, the model architecture and the hyper\hyp{parameters}. However, this only occurs at the beginning of the learning process.

\subsection{\sysname Overview}
\label{subsec:overview}

\begin{figure}[h]
    \centering
    \includegraphics[width=\linewidth]{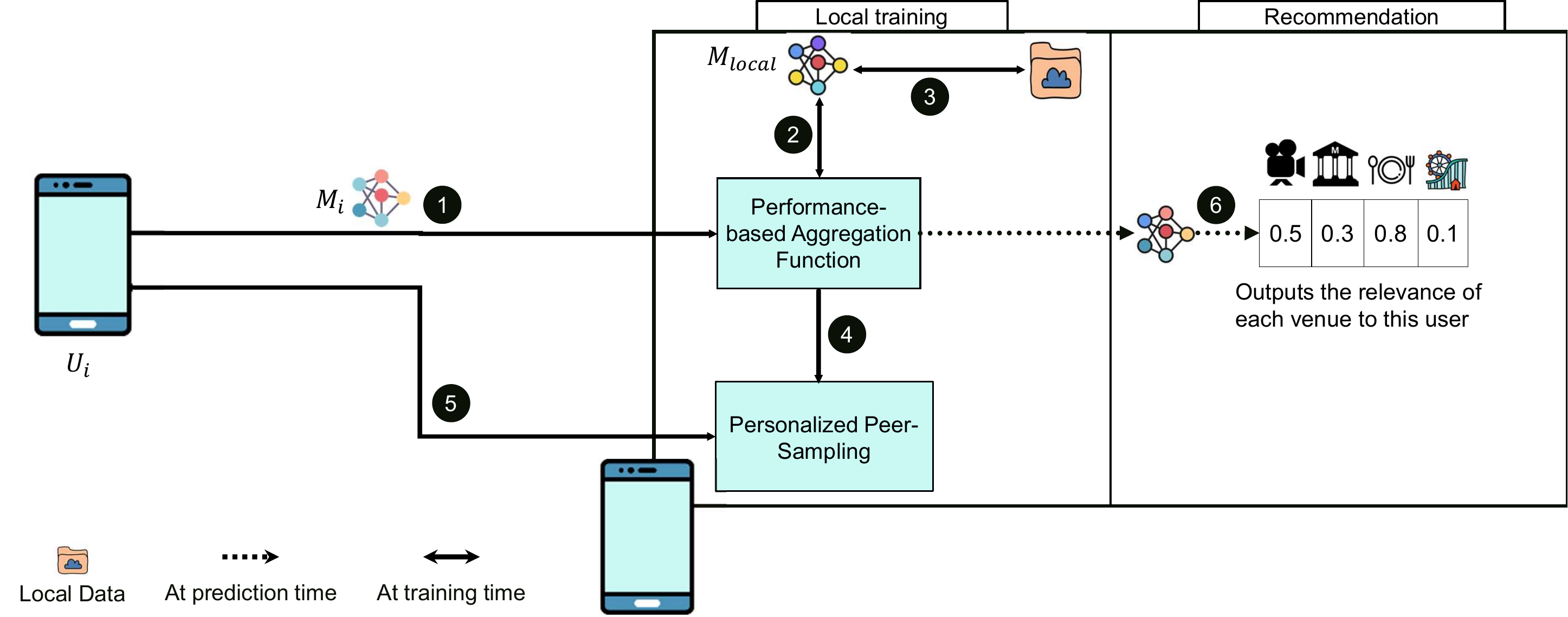}
    \caption{\sysname System Main Components.}
    \label{fig:SystemOverview}
\end{figure}

\noindent Figure \ref{fig:SystemOverview} presents an overview of \sysname, which operates in two main phases: a training phase and a recommendation phase.
The training phase makes use of two main components depicted in the left part of the figure: a performance-based model aggregation function and a personalized peer-sampling protocol. Upon receiving a model, say $M_i$ from one of its neighbors, say $U_i$ (step \cercle{1}), the performance-based model aggregation function assesses the local performance of $M_i$ and aggregates the latter with its latest version of the local model $M_{local}$ (step \cercle{2}). Periodically, this model is further trained with new local data (e.g., new ratings) as depicted in step \cercle{3} of the figure. The information about the performance of $M_i$ is not only used to adjust the way the latter is aggregated with $M_{local}$, it is also sent to the personalized peer\hyp{sampling} protocol (step \cercle{4}). The objective of this protocol is to periodically update nodes' views (step \cercle{5}). At the very beginning, view updates are performed randomly to leverage previous theoretical and empirical works~\cite{Bollobs2004TheDO,ThePeerSamplingService:ExperimentalEvaluationofUnstructuredGossip-BasedImplementations} on how to ensure a logarithmic time dissemination of models. Later, when the received models start to exhibit good local performance, the personalized peer-sampling service remembers good neighbors and uses this information to improve the quality of the view. In \sysname, both the performance-based model aggregation function and the personalized peer sampling protocol are totally decentralized protocols that do not rely on any central entity in the system.

\section{\sysname Detailed Description}
\label{sec:details}
In this section, we present the two key components of \sysname, \ie, the personalized peer-sampling protocol and the performance-based model aggregation function.

\subsection{Performance\hyp{based Aggregation Function}}
\label{subsec:performancebasedaggregation}
\begin{algorithm}[!htb]
\DontPrintSemicolon
\KwData{Local Dataset $D_i$, currentModel $M_i$, $WeightingSize$, $TestSize$}

\SetKwFunction{Init}{Init}
\SetKwFunction{PerformanceBasedAggregationFunction}{PerformanceBasedAggregationFunction}
\SetKwProg{Fn}{Procedure}{:}{}
\Fn{\Init}{

$D_i^{Weighting} = RandomlySamplesSet(D_i, WeightingSize)$ \Comment{Randomly samples the weighting set}
\label{lst:line2:1}

$D_i^{test} = RandomlySamplesSet(D_i \setminus D_i^{Weighting}, TestSize)$
\Comment{Randomly samples the test set}

$D_i^{train} = D_i \setminus (D_i^{Weighting} \bigcup D_i^{test}) $

} \;

\Fn{\PerformanceBasedAggregationFunction{$M_x$}}{
$ P_x = PredictAndEvaluate(M_x, D_i^{weighting})\;$ \Comment{Make prediction with $M_x$ and evaluate its performance}  \label{lst:line2:2}

$ P_i = PredictAndEvaluate(M_i, D_i^{weighting})\;$ \Comment{Make prediction with $M_i$ and evaluate its performance} \label{lst:line2:4}

$M_i= \frac{1}{P_i + P_x} (P_i \times M_i + P_x \times M_x) $ \Comment{Aggregate $M_i$ and $M_x$ w.r.t their performance} \label{lst:line2:6}

$Update(M_i, D_i^{train})$ \Comment{Train the new $M_i$ on local data} \label{lst:line2:7}

  }
  \;

\caption{Performance\hyp{based} Aggregation Function}
\label{alg:performancebased}
\end{algorithm}

\noindent As opposed to existing model aggregation functions, which aim at optimizing global performance, we propose a method that aims at maximizing the local performance of nodes. By doing so, nodes are more likely to obtain satisfactory recommendations, which can improve both average and tail performance. To reach this objective, a node needs to (i) assess the quality of a model received from one of its neighbors and (ii) use this information during the model aggregation phase.

Let us consider the gossip learning setup described in section~\ref{sec:background}, where each node $C_i$ has its own local data $D_i$ and maintains a local model, say $M_i$. In order to assess the quality of received models, $C_i$ uses a random subset $D_i^{weighting}$ taken from its local data set $D_i$, which we refer to as the weighting set of $C_i$ (Algorithm~\ref{alg:performancebased}, line~\ref{lst:line2:1}).
$D_i^{weighting}$ can be seen as a validation set so it cannot be used for model training.
However, in scenarios where the data points are produced in a continuous stream, $D_i^{weighting}$ can be refreshed with new data points while some older points from $D_i^{weighting}$ can be appended to the training set. 
Upon reception of a model $M_x$ from one of its neighbors $C_x$, $C_i$ evaluates it on $D_i^{weighting}$. More specifically, $M_x$ is required to provide $C_i$ with a top-K recommendation list for each item $\in D_i^{weighting}$ among 100 random items (see Section \ref{subsec:methodology}). By quantifying how high items of $D_i^{weighting}$ are ranked, a measure of performance of $M_x$, say $P_x$ is obtained (Algorithm~\ref{alg:performancebased}, line~\ref{lst:line2:2}). Intuitively, this measure quantifies how similar the data distribution $M_x$ was trained on, is to the distribution of $D_i$. In the context of recommender systems, a node's best\hyp{received} models are more likely to come from nodes who have similar tastes. Then, in order to use this information during the model aggregation phase, the node proceeds by computing the performance of its local model $M_i$ on $D_i^{weighting}$, say $P_i$ (Algorithm~\ref{alg:performancebased}, line \ref{lst:line2:4}). This will allow $C_i$ to compare the two models and assess the potential contribution of each. As illustrated in line~\ref{lst:line2:6} of Algorithm \ref{alg:performancebased}, $P_i$ and $P_x$ are used as weights by the model aggregation function. By doing so, nodes give more substance to the models that perform better on their data. 
This pairwise aggregation principle is classical in gossip learning works~\cite{Decentralizedlearningworks:Anempiricalcomparisonofgossiplearningandfederatedlearning,DecentralizedRecommendationBasedonMatrixFactorization:AComparisonofGossipandFederatedLearning} where nodes can have substantially different model pushing periods, that is, a node rarely receives multiple models at the same time. Therefore, creating synchronicity in the network by waiting for a set of models can hinder convergence. Hence, by following the pairwise principle, \sysname reduces the synchronicity, enabling models to be aggregated as soon as they are received.

Following this, a traditional model update is done on the train data, that is, $D_i^{train}$~(Algorithm~\ref{alg:performancebased}, line \ref{lst:line2:7}). At last, $C_i$ stores $P_x$ as the last model performance received from $C_x$. This information will be used by the personalized peer\hyp{sampling} protocol as further discussed in the following section.


\subsection{Personalized Peer\hyp{Sampling} Algorithm}\label{personalized_peersampling}
\begin{figure}[h]
    \centering
    \includegraphics[width=\linewidth]{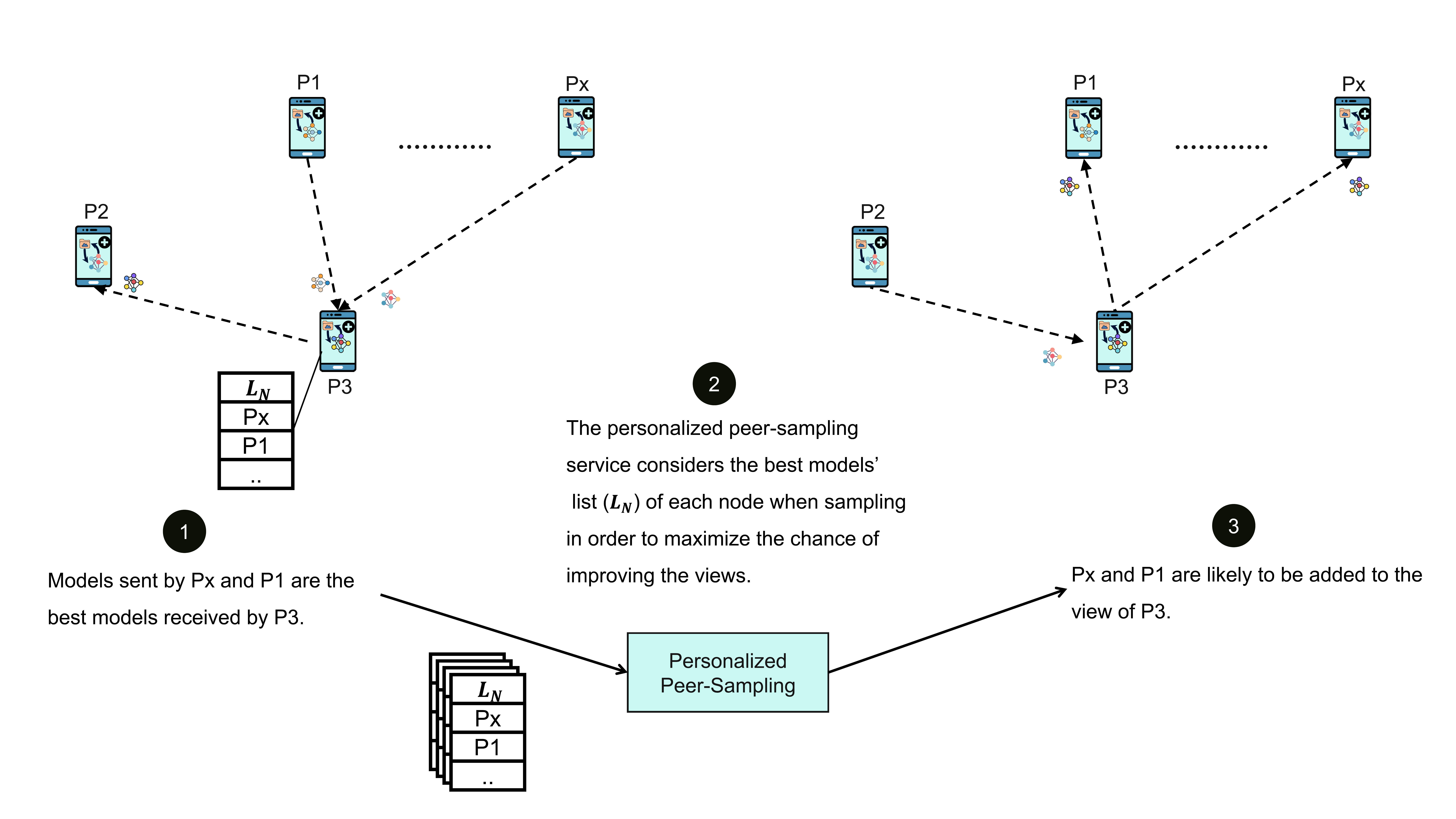}
    \caption{Personalized Peer\hyp{Sampling} Execution Scenario.}
    \label{fig:peer_sampling}
\end{figure}
\noindent As discussed in Section~\ref{sec:background}, peer\hyp{sampling} can have a significant impact on the convergence speed of Gossip Learning algorithms. This protocol is even more important in use cases where a node might be more interested in models coming from similar nodes in terms of data distribution. 
To take this into account, we extend the random peer\hyp{sampling} protocol by considering the quality of the current nodes in the view, that is, their ability of sending quality models, as illustrated in Figure~\ref{fig:peer_sampling}. To this end, the personalized peer-sampling service collects the recommendation performance of the received models and the identity of their respective nodes (steps \cercle{1} and \cercle{2} in the figure). These performances come from the evaluation step of the performance-based aggregation function. As a result, nodes that sent quality models to a given node are likely to join his view in later rounds (step \cercle{3} in the figure).

More formally, let us consider a node $C_i$ during a peer\hyp{sampling} phase.
To update its view $V_{C_i}$, $C_i$ performs two main steps, as illustrated in Algorithm \ref{alg:peersampling}. First, it relies on past information collected from the performance-based model aggregation function regarding the performance of models received from other nodes in the system. Specifically, $C_i$ maintain a list $L_N$ of nodes from which it received models in the past, sorted by the performances of these models (lines~\ref{lst:line3:1}-\ref{lst:line3:2}). To update its view, $C_i$ considers the top $T$ elements of $L_N$ (lines~\ref{lst:line3:3}-\ref{lst:line3:4}). 
In addition to this first step, $C_i$ randomly collects $R$ nodes from its peers' views to form the new view (lines~\ref{lst:line3:5} -\ref{lst:line3:7}). $T$ and $R$ depend on an exploration/exploitation ratio $\alpha$. Values of $\alpha$ close to zero translate to an exploitation\hyp{dominant} approach that maximizes the performance of the view at the expense of discovering the network (\ie, $P'$ $\approx$ $T$) whereas on the contrary $\alpha$ = 1 is an exploration\hyp{dominant} approach and is equivalent to the traditional random peer\hyp{sampling} (\ie, $P'$ = $R$). 
The optimal value of $\alpha$ cannot really be determined as it is highly dependent on the underlying data. However, we observed empirically that an $\alpha = 0.4$, equivalent to 40\% of nodes coming from $L_N$ and 60\% of random nodes, gives the best results in our experiments.

\begin{algorithm}[!htb]
\DontPrintSemicolon
\KwData{Nodes set $N$, currentModel $M_i$, exploitation/exploration ratio $\alpha$, list of received models and their best performance $L_N$, $V_{C_{i}}$}

\SetKwFunction{Get}{Get}
\SetKwFunction{UpdateView}{UpdateView}
\SetKwFunction{Size}{Size}
\SetKwFunction{PerformanceBasedAggregationFunction}{PerformanceBasedAggregationFunction}
\SetKwFunction{OnModelReceived}{OnModelReceived}
\SetKwFunction{AddSorted}{AddSorted}
\SetKwProg{Fn}{Procedure}{:}{}
\Fn{\OnModelReceived{$M_x$}}{

$P_x$ = \PerformanceBasedAggregationFunction($M_x$) \label{lst:line3:1}

$L_N.\AddSorted(C_x,P_x)$ \Comment{Add received model's performance and its owner to the sorted list}
\label{lst:line3:2}
} \;

\Fn{\UpdateView{}}{
$ T = (1 - \alpha) \times \lvert{V_{C_{i}}}\rvert$
\label{lst:line3:3}

$ExploitationPeers = L_N.\Get(T) $ \Comment{Get the top T elements of $L_N$ performance wise}
\label{lst:line3:4}

$ R =  \lvert{V_{C_{i}}}\rvert - T $
\label{lst:line3:5}

$ExplorationPeers = RandomPeers(N,R)$ \Comment{Get R random peers from N}
\label{lst:line3:6}

$V_{C_{i}} = Concat(ExplorationPeers,ExploitationPeers) $ \Comment{Update view with exploitation and exploration peers}\label{lst:line3:7}
  }
  \;

\caption{Personalized Peer\hyp{Sampling} Algorithm}
\label{alg:peersampling}
\end{algorithm}

\section{Performance Evaluation}
\label{sec:eval}
In this section, we evaluate \sysname on three real world datasets and compare its performance against centralized and decentralized solutions. More specifically, we aim at answering the following research questions:

\begin{itemize}
\item \textbf{RQ1:} \sysname aggregates models based on their local performance. How efficient is \sysname compared to decentralized solutions (that aggregate models to target average performance) and can it be competitive with federated baselines?\vspace{0.2cm} 
\item \textbf{RQ2:} Does considering the local performance of models make them more tailored to their users and how does that affect tail performance?\vspace{0.2cm}
\item \textbf{RQ3:} What role does the amount of sparsity play in the context of an increased personalization?\vspace{0.2cm}
\item \textbf{RQ4:} \sysname extends the random\hyp{peer}\hyp{sampling} protocol by considering the local model performance. How does this affect the learning process?\vspace{0.2cm}
\item \textbf{RQ5:} How sensitive is \sysname to the peer set size parameter?\vspace{0.2cm}
\item \textbf{RQ6:} What level of overhead does \sysname incur on the recommendation pipeline?\vspace{0.2cm}

\end{itemize}

The rest of this section is structured as follows. 
We first introduce the experimental environment we used to evaluate \sysname (Section~\ref{subsec:envsetup}) before describing the used datasets (Section~\ref{subsec:datasets}). 
Further, we explain our evaluation methodology and the employed evaluation metrics (Section~\ref{subsec:methodology}). 
Finally, we present our competitors (Section~\ref{subsec:baseline}) and analyze our experimental results answering the six questions introduced above (Section~\ref{subsec:results}).

\subsection{Experimental Environment}\label{subsec:envsetup}

\sysname is built on top of a decentralized network of nodes that is simulated using \emph{OMNetPy}~\cite{omnetpy}, a python interface for \emph{OMNet++}~\cite{omnetpp}, a popular simulation platform for communication networks. 
Associated with the well known \emph{TensorFlow/Keras}~\cite{tensorflow}, it allows training models at the level of each node and exchanging these models over a decentralized network.
It is worth noting that since \emph{OMNet++} is an event\hyp{based} simulator and our experiments are based on round\hyp{based} behaviors, we use timed events to simulate these actions. 
Note that in our simulations initial nodes' neighbors were generated randomly. Nevertheless, to ensure reliability and put all decentralized competitors in the same setup, experiments were repeated several times using the same seed when generating the initial setup.

\subsection{Datasets}\label{subsec:datasets}
\begin{table}[!htb]
\parbox{\linewidth}{
    \centering
    \resizebox{0.8\textwidth}{!}{
   \begin{tabular}{|| c|c|c|c|c|c|c ||}
         \hline
         Dataset & Type & Users & Locations/Movies & Records & Sparsity \% \\
         \hline\hline
         Foursquare\hyp{NYC} &  Points of interest & 1083 & 38333 & 227,428 & 
         0.997
        \\
         \hline
         Gowalla\hyp{NYC} & Points of interest & 718 & 32924 & 185,932 & 0.986
        \\
         \hline
         MovieLens\hyp{100k} &  Movies Recommendation & 943 & 1682 & 100,000 & 0.936
        \\
        \hline
    \end{tabular}
}    
    \caption{Statistics of Datasets.}
     \label{tab:datasets}
}
\end{table}
\noindent As presented in Table~\ref{tab:datasets},
for our experiments we chose three different datasets that are commonly used to evaluate recommendation systems. 
MovieLens\hyp{100K}~\cite{movielens} is a dataset containing 100k movie ratings from 943 users on 1682 movies. 
Similarly to previous works~\cite{muhammad2020fedfast,Fast-adaptingandPrivacy-preservingFederatedRecommender,Fast-adaptingandPrivacy-preservingFederatedRecommender} we adapted this dataset to the Generalized Matrix Factorization (GMF) as follows: (i) all user ratings are converted to positive ratings (\ie, 1) and (ii) items not rated by a user are labeled 
with zero. This pre\hyp{processing} is closely related to the fact that the GMF model is a classifier whose output is subject to a probabilistic activation function (\ie, the sigmoid function) representing the probability of relevance of an item to a user~\cite{NCF}. Furthermore,  
the work of~\citet{koren2008factorization} details how incorporating such binary data, which normalizes the interpretation of ratings by users, can improve prediction accuracy. Finally, we note that our competitors~\cite{muhammad2020fedfast,Fast-adaptingandPrivacy-preservingFederatedRecommender} adopted the same approach.

The two other datasets are Foursquare\hyp{NYC} and Gowalla\hyp{NYC} Check-ins. The former consists of 227,428 check-ins collected by 1083 users from 38333 venues and the latter is composed of 82197 check-ins and 718 users, both in New York city.
Each check-in is associated with its time stamp, its GPS coordinates and the venue category.
In the pre-processing phase, we removed locations with less than 10 visitors, and users with less than 10 check-ins, as it is common for Point-of-Interest recommender systems~\cite{Guo21,Qiang19}.
In addition, similarly to  MovieLens, when evaluating GMF, we transform the check-ins to binary ratings in order to build a top-K ranking recommender system. This step is not necessary for evaluating the  PRME-G model that is specifically designed for sequential check\hyp{in} data.
We have also computed the sparsity of each dataset as in Equation~\eqref{eq:sparsity}. It can be observed that MovieLens has significantly less sparsity than the other two datasets.

\begin{equation}\label{eq:sparsity}
    sparsity = 1 - \frac{avg.~rated~items}{all~items} 
\end{equation}

\subsection{Evaluation Methodology}
\label{subsec:methodology}
To evaluate \sysname, we split the local datasets of each user using the 85-15 rule: 85\% of the data is used to train the model, which will be evaluated on the remaining 15\%.
Specifically, for each client $C_i$ , we split its data into a training set $D_i^{train}$ and a test set $D_i^{test}$.
As explained in Section~\ref{subsec:performancebasedaggregation}, for \sysname, a weighting set $D_i^{weighting}$ with the same size as the test set is furthermore extracted from $D_i^{train}$. 
The model $M_i$ is trained on $D_i^{train}$ and locally evaluated on $D_i^{test}$ as opposed to other works which evaluate models on the union of test sets (\ie, a global test set). By doing so, we measure the local performance of each model and quantify the user's satisfaction. 
More specifically, for a given user, we require each algorithm to provide the top-K
recommendation list for each test item among 100 randomly sampled and unvisited/unrated items. This is a common strategy to avoid ranking test items among all unseen items~\cite{surveyrs,NCF}, which can be considerably time-consuming. The value of K is set to 5, 10 and 20 respectively.

For the GMF model, we use two classic metrics to evaluate the ranked lists, namely the hit ratio at rank r (\ie, HR@r), which is the performance metric we use to weight models in our aggregation function, and the normalized discounted cumulative gain at rank r (\ie, NDCG@r). 
Intuitively, HR@r of the item $t$ takes the value 1 if the latter is present in the top r elements of the ranked list and 0 otherwise. 
NDCG@r measures how good is the position of $t$ in the list, assigning higher scores to hits at the top of the ranked list. More specifically, HR@r of a client $C_i$ will be computed as in Equation \eqref{eq:HRglobal}, while its NDCG@r will be the average of NDCGs@r of each of its items, computed each as in Equation \eqref{eq:NDCGitem}.

\begin{equation}\label{eq:HRglobal}
    HR@r = \frac{\sum_{j=0}^{I} HR@r_j}{\lvert{D_i^{test}}\rvert} 
\end{equation}
where $HR@r_j$ is the hit ratio for item j and $I$ the set of items of the user $C_i$.

\begin{equation}\label{eq:NDCGitem}
    NDCG@r_i = \frac{\log(2)}{\log(p+2)} 
\end{equation}
  with $i$ being the corresponding item and $p$ its position in the ranked list.

For the PRME-G model, we resort to the widely known metrics used in the original paper of~\citet{feng2015personalized}, namely Precision@r and Recall@r, which we use to compute the F1\hyp{Score} at rank r. These metrics are well suited for a next POI recommendation use case, where the ability of a model to find the known relevant POIs for a user (\ie, recall) and its capacity to distinguish between relevant and non-relevant locations (\ie, precision) is very important.
More precisely, Precision@r is the proportion of recommended items in the top\hyp{r} set that are relevant. 
Hence, it is computed by taking the ratio of relevant items found in the top\hyp{r} divided by r (see Equation~\eqref{eq:PrecsionK}). 
Similarly, Recall@r is the proportion of relevant items found in the top\hyp{r} recommendations.
It is computed by taking the ratio of relevant items found in the top r divided by the total number of relevant items for a particular user (see Equation~\eqref{eq:RecallK}). 
To combine these two metrics, we compute the F1\hyp{Score}@r and use it both for evaluation and aggregation purposes. F1\hyp{Score}@r metric takes into account both Precision@r and Recall@r, as showed in Equation~\eqref{eq:F1ScoreK}.

\begin{equation}\label{eq:PrecsionK}
    Precision@r_u = \frac{\lvert{\textnormal{Relevant items for user u@r}}\rvert}{r} 
\end{equation}

\begin{equation}\label{eq:RecallK}
    Recall@r_u = \frac{\lvert{\textnormal{Relevant items for user u@r}}\rvert}{\textnormal{All relevant items for user u}} 
\end{equation}

\begin{equation}\label{eq:F1ScoreK}
    F1\textnormal{-}Score@r_u = 2 \times \frac{Precision@r_u \times Recall@r_u}{Precision@r_u + Recall@r_u} 
\end{equation}

\subsection{Baselines}\label{subsec:baseline}
We compare our aggregation techniques against six baselines. These baselines are of two kinds: federated baselines, \ie, baselines that follow the classical FL architecture and decentralized baselines, \ie, baselines that rely on Gossip Learning.
\subsubsection{Federated Baselines}
\begin{itemize}
\item \emph{FedAvg}~\cite{FederatedLearningofDeepNetworksusingModelAveraging} is one of the most popular FL protocols, based on the traditional master-slave FL architecture. Instead of centralizing the raw data as in Centralized GMF, FedAvg centralizes model updates from the FL users and aggregates them based on the number of samples they were trained on.\vspace{0.2cm}
{\item  \emph{FedFast}~\citep{muhammad2020fedfast} is the closest work to ours, albeit being centralized. 
FedFast is an extension over the FedAvg algorithm that clusters users at each round using a k\hyp{means} algorithm. 
Subsequently, it performs cross\hyp{cluster} aggregation for the user embeddings of the GMF model while aggregating the item embeddings and other model parameters in the same manner as FedAvg. At the beginning of each FL round, it samples users equally from each cluster. We adapt this technique for the PRME-G model in order to compare it with \sysname.}\vspace{0.2cm}
{ \item \emph{Reptile}~\citep{Fast-adaptingandPrivacy-preservingFederatedRecommender} is a meta\hyp{learning} approach, which uses the weights learned by the FedAvg algorithm to initialize a local meta\hyp{model} to each user. Afterwards, this meta-model is fine\hyp{tuned} on local data by updating the initial parameters towards the final parameters learned locally. This approach has been shown to improve personalization.}
\end{itemize}
\subsubsection{Decentralized Baselines}
\begin{itemize}
\item \emph{Model\hyp{Age}\hyp{Based}} is a decentralized gossip\hyp{based} approach introduced in \cite{Decentralizedlearningworks:Anempiricalcomparisonofgossiplearningandfederatedlearning,DecentralizedRecommendationBasedonMatrixFactorization:AComparisonofGossipandFederatedLearning}.
Models received by a user are weighted proportionally to the number of times they have been updated (\ie, the model's age).
The rationale behind this aggregation algorithm is that a model which was updated by a larger number of nodes has more information so it should weight more in the aggregation function.\vspace{0.2cm}
\item \emph{Decentralized FedAvg}~\citep{fullavg} is a decentralized version of FedAvg.
Models circulate freely through the network and nodes weight them in the aggregation function depending on the number of samples they were trained on.\vspace{0.2cm}
\item \emph{Decentralized Reptile} is a decentralized version of Reptile~\citep{Fast-adaptingandPrivacy-preservingFederatedRecommender} that we implemented in which each node fine\hyp{tunes} its local model based on the parameters learned in a decentralized collaborative way.
\end{itemize}

\subsection{Experimental Results and Discussions}\label{subsec:results}

Based on the evaluation setup described above, we conduct experiments to evaluate the performance of \sysname against the state-of-the-art solutions introduced in the previous section. Our evaluation aims at answering the questions introduced at the beginning of Section~\ref{sec:eval}.

\subsubsection{\sysname Average Performance Comparison}
\label{subsec:averageperf}
\begin{table}[!htb]
\parbox{.47\linewidth}{
 \captionsetup{justification=centering}

    \centering
    \resizebox{0.46\textwidth}{!}{
   \begin{tabular}{|| c|c|c|c ||}
         \hline
         \multirow{2}{2cm}{Algorithm} & \multicolumn{3}{c|}{Average Hit Ratio \%}\\
         \cline{2-4}
         & K = 20 & K = 10 & K = 5\\
         \hline
         FedAvg & 32.42 & 24.0 & 18.30 \\
         \hline
         Reptile &  31.59 & 23.25 & 18.10
         \\
         \hline
         FedFast &  37.12 & 26.27 & 18.94
         \\
         \hline
         Decentralized FedAvg & 37.00 & 28.10 & 20.44 \\
         \hline
         Model\hyp{Age}\hyp{Based} & 38.54 & 26.87 & 20.16
         \\
         \hline
         Decentralized Reptile &  38.58 & 28.99 & 22.70
         \\
         \hline
         \textbf{\sysname} &  \textbf{45.93} & \textbf{32.59} & \textbf{25.90}
         \\
         \hline
        
    \end{tabular}
}   
    \caption{{Average Hit Ratio \% on Foursquare-NYC \\(GMF).}}
     \label{tab:averagehrfoursquare}
}
\parbox{.47\linewidth}{
\captionsetup{justification=centering}

\centering
\resizebox{0.46\textwidth}{!}{
\begin{tabular}{|| c|c|c|c ||}
         \hline
         \multirow{2}{2cm}{Algorithm} & \multicolumn{3}{c|}{Average NDCG \%}\\
         \cline{2-4}
         & K = 20 & K = 10 & K = 5\\
         \hline
        FedAvg & 16.61 & 14.37 & 12.66 
         \\
         \hline
        Reptile &  17.12 & 14.36 & 13.38
         \\
         \hline
        FedFast &  18.41 & 15.69 & 13.30
         \\
         \hline
        Model\hyp{Age}\hyp{Based} & 19.07 & 16.81 & 14.92
         \\
         \hline
        Decentralized FedAvg & 18.75 & 17.39 & 15.37
         \\
         \hline
        Decentralized Reptile &  22.04 & 19.61 & 17.63
         \\
         \hline
      
        \textbf{\sysname} & \textbf{23.10} & \textbf{20.44} & \textbf{19.24}
        \\
         \hline
        
    \end{tabular}
} 
    \caption{{Average NDCG \% on Foursquare-NYC \\(GMF).}}
    \label{tab:averagendcgfoursquare}
}
\end{table}
\begin{table}[!htb]
\parbox{.47\linewidth}{
\captionsetup{justification=centering}

    \centering
    \resizebox{0.46\textwidth}{!}{
   \begin{tabular}{|| c|c|c|c ||}
         \hline
         \multirow{2}{2cm}{Algorithm} & \multicolumn{3}{c|}{Average Hit Ratio \%}\\
         \cline{2-4}
         & K = 20 & K = 10 & K = 5\\
         \hline
         
         FedAvg & 79.69 & 64.4 & 47.59 \\
         \hline
         Reptile & 79.49 & 62.6 & 44.1
         \\
         \hline
         FedFast &  \textbf{82.79} & \textbf{67.80} & \textbf{50.00}
        \\
         \hline
         Decentralized FedAvg & 74.9 & 55.0 & 40.7 \\
         \hline
         Model\hyp{Age}\hyp{Based} & 73.69 & 55.5 & 42.4
         \\
         \hline
         Decentralized Reptile & 70.19 & 51.60 & 36.29
         \\
         \hline
         \textbf{\sysname} &  79.29 & 61.20 & 47.59
        \\
         \hline
        
    \end{tabular}
 }   
    \caption{{Average Hit Ratio \% on MovieLens\hyp{100k} \\(GMF).}}
     \label{tab:averagehrmovielens}
}
\parbox{.47\linewidth}{
\captionsetup{justification=centering}

    \centering
    \resizebox{0.46\textwidth}{!}{
       \begin{tabular}{|| c|c|c|c ||}
         \hline
         \multirow{2}{2cm}{Algorithm} & \multicolumn{3}{c|}{Average NDCG \%}\\
         \cline{2-4}
         & K = 20 & K = 10 & K = 5\\
         \hline
         FedAvg & 41.03 & 37.35 & 32.07 \\
         \hline
         Reptile & 40.07 & 35.80 & 29.83 \\
         \hline
         FedFast & \textbf{44.14} & \textbf{40.34} & \textbf{34.50} \\
         \hline
         Decentralized FedAvg & 36.17 & 31.79 & 27.56 \\
         \hline
         Model\hyp{Age}\hyp{Based} & 37.47 & 33.06 & 29.81 \\
         \hline     
         Decentralized Reptile & 34.0 & 29.05 & 24.14 \\
         \hline
         \textbf{\sysname} & 40.09 & 35.60 & 32.53 \\
         \hline
        
    \end{tabular}
    }
    \caption{{Average NDCG \% on MovieLens\hyp{100k} \\(GMF).}} 
    \label{tab:averagendcgmovielens}
    }
\end{table}
\begin{table}[!htb]
\parbox{.47\linewidth}{
\captionsetup{justification=centering}

    \centering
    \resizebox{0.46\textwidth}{!}{
       \begin{tabular}{|| c|c|c|c ||}
         \hline
         \multirow{2}{2cm}{Algorithm} & \multicolumn{3}{c|}{Average F1-Score \%}\\
         \cline{2-4}
         & K = 20 & K = 10 & K = 5\\
         \hline
         FedAvg & 32.87 & 28.62 & 21.50 \\
         \hline
         Reptile & \textbf{35.00} & 31.53 & 23.86 \\
         \hline
         FedFast & 33.35 & 30.56 & 23.18 \\
         \hline
         Decentralized FedAvg & 21.82 & 18.70 & 16.03 \\
         \hline
         Model\hyp{Age}\hyp{Based} & 20.48 & 18.60 & 17.19 \\
         \hline     
         Decentralized Reptile & 27.28 & 25.46 & 21.19 \\
         \hline
         \textbf{\sysname} & 34.05 & \textbf{32.44} & \textbf{26.38} \\
         \hline
        
    \end{tabular}
}   
\caption{Average F1-Score \% on Foursquare\hyp{NYC} (PRME-G).}
    \label{tab:averagef1foursquare}
    }
\parbox{.47\linewidth}{
 \captionsetup{justification=centering}
  
    \centering
    \resizebox{0.46\textwidth}{!}{
    \begin{tabular}{|| c|c|c|c ||}
     \hline
     \multirow{2}{2cm}{Algorithm} & \multicolumn{3}{c|}{Average F1-Score \%}\\
     \cline{2-4}
     & K = 20 & K = 10 & K = 5\\
     \hline
     FedAvg & 21.35 & 18.53 & 15.37 \\
     \hline
     Reptile & 20.90 & 17.07 & 13.46 \\
     \hline
     FedFast & 21.80 & 18.28 & 15.09 \\
     \hline
     Decentralized FedAvg & 11.30 & 10.18 & 9.13 \\
     \hline
     Model\hyp{Age}\hyp{Based} & 9.92 & 8.59 & 7.64 \\
     \hline     
     Decentralized Reptile & 9.10 & 7.66 & 6.05 \\
     \hline
     \textbf{\sysname} & \textbf{23.63} & \textbf{20.66} & \textbf{17.15} \\
     \hline
    
\end{tabular}
}
  \caption{Average F1-Score \% on Gowalla\hyp{NYC} \\(PRME-G).} 
    \label{tab:averagef1gowalla}
    }
\end{table}
In this experiment, we compare the performance of all models on \sysname and the aforementioned baselines.
Firstly, we evaluate the \emph{average} top-K recommendation quality where K is 5, 10 and 20.  
Table~\ref{tab:averagehrfoursquare} and Table~\ref{tab:averagehrmovielens} show the average
top-20 recommendation hit ratio comparison on Foursquare and MovieLens, while Table~\ref{tab:averagendcgfoursquare} and Table~\ref{tab:averagendcgmovielens}
show the average NDCG. 
{In Tables~\ref{tab:averagef1foursquare} and~\ref{tab:averagef1gowalla} we report the F1 score of PRME-G on the Foursquare and Gowalla datasets.
From these results, we can observe that \sysname outperforms its decentralized competitors on all datasets.}
Compared to Decentralized FedAvg, \sysname has a better HR@20 and NDCG@20 by a margin of 8.93\% and 4.35\%, respectively.
The results for different values of K and two different metrics confirm that \sysname does not privilege recall to the ranking quality of tested items, even though HR is used during the aggregation. 
This is essential to show that our system is not only optimizing a specific metric (\ie, Goodhart's law~\cite{manheim2018categorizing}) but using it to optimize the true quality of the model. 

Foursquare is the most sparse dataset out of the three (see Table~\ref{tab:datasets}) and therefore has more heterogeneous user profiles so one global model may not be able to fit all user preferences (see §\ref{subsec:sparsity}). Therefore, our solution which builds personalized models outperforms centralized solutions in most cases except for the F1@20 on Foursquare (see Table~\ref{tab:averagef1foursquare}). The sparsity observation is further supported by the results obtained on Gowalla (see Table~\ref{tab:averagef1gowalla}), which is similar in sparsity to Foursquare and where \sysname outperforms all the other solutions, for all values of K.
The Reptile algorithm has a fairly high performance with the PRME-G model and is the best algorithm for K=20 on Foursquare. However, since it is outperformed by \sysname for other values of K, we can conclude that it finds many of the relevant items but does not rank them as accurately as \sysname. On MovieLens, which is less sparse (see Tables \ref{tab:averagehrmovielens} and \ref{tab:averagendcgmovielens}), \sysname is usually overpassed by centralized solutions (\ie, FedFast, FedAvg and Reptile).
Among the latter three, FedFast performs the best since it personalizes based on a view over all users so it reaches a better personalization-generalization tradeoff compared to \sysname. Nevertheless, \sysname still substantially overpasses all the decentralised algorithms. 

Based on these results, we can observe that \textbf{\sysname always outperforms its decentralized competitors on average performance} and can even be competitive with centralized solutions, especially in the case of sparse recommendation matrices~\textbf{(RQ1)}.

\subsubsection{\sysname Tail Performance Comparison}
\begin{figure}[!htb]
\centering
\begin{subfigure}[b]{0.45\textwidth}
    \centering
    \includegraphics[scale=0.5]{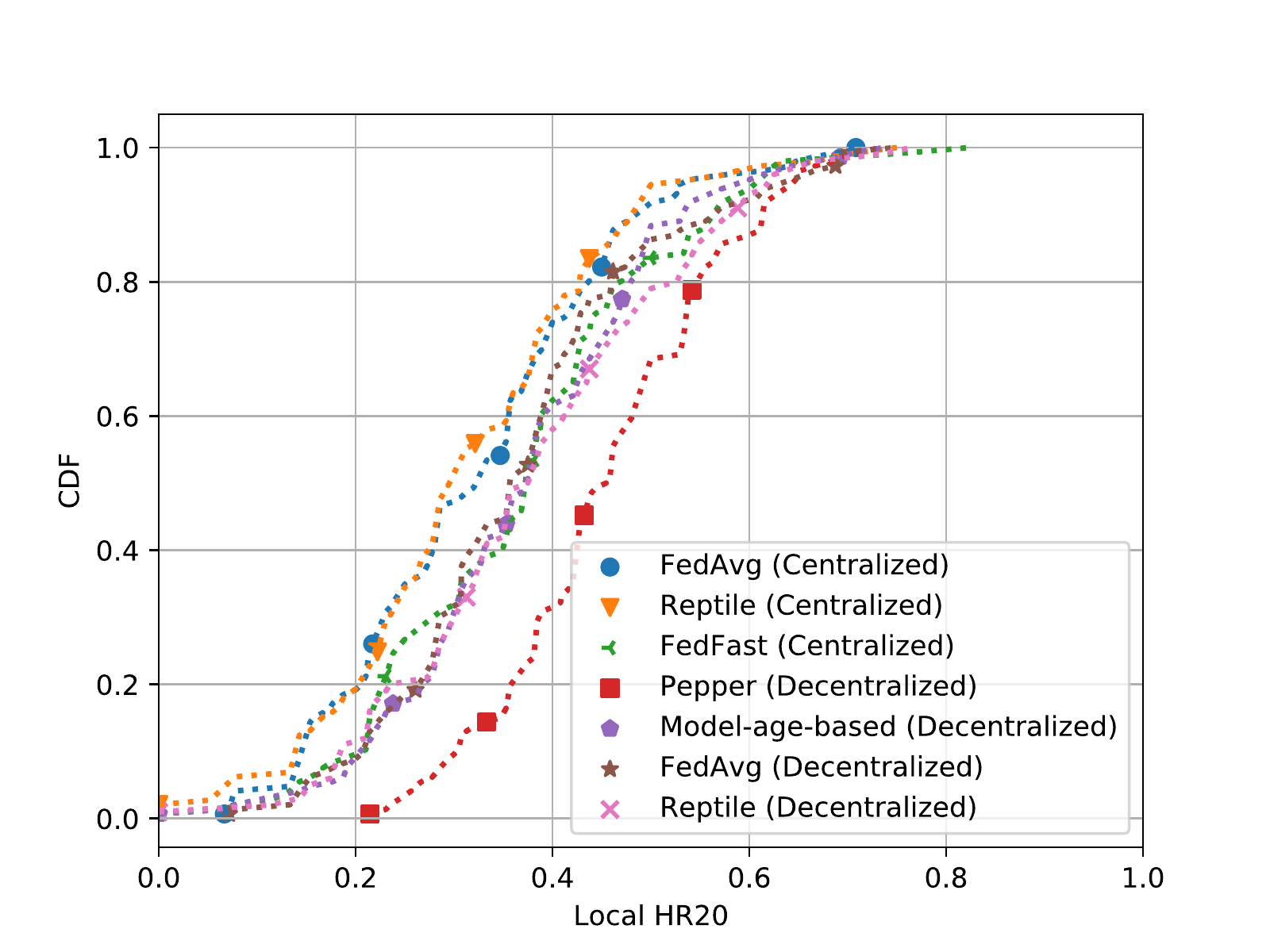}
    \caption{{Local Hit Ratio@20 cumulative distribution function.}}%
     \label{fig:percentilehrfoursquare}
\end{subfigure}
\hfill
\begin{subfigure}[b]{0.45\textwidth}
    \centering
    \includegraphics[scale=0.5]{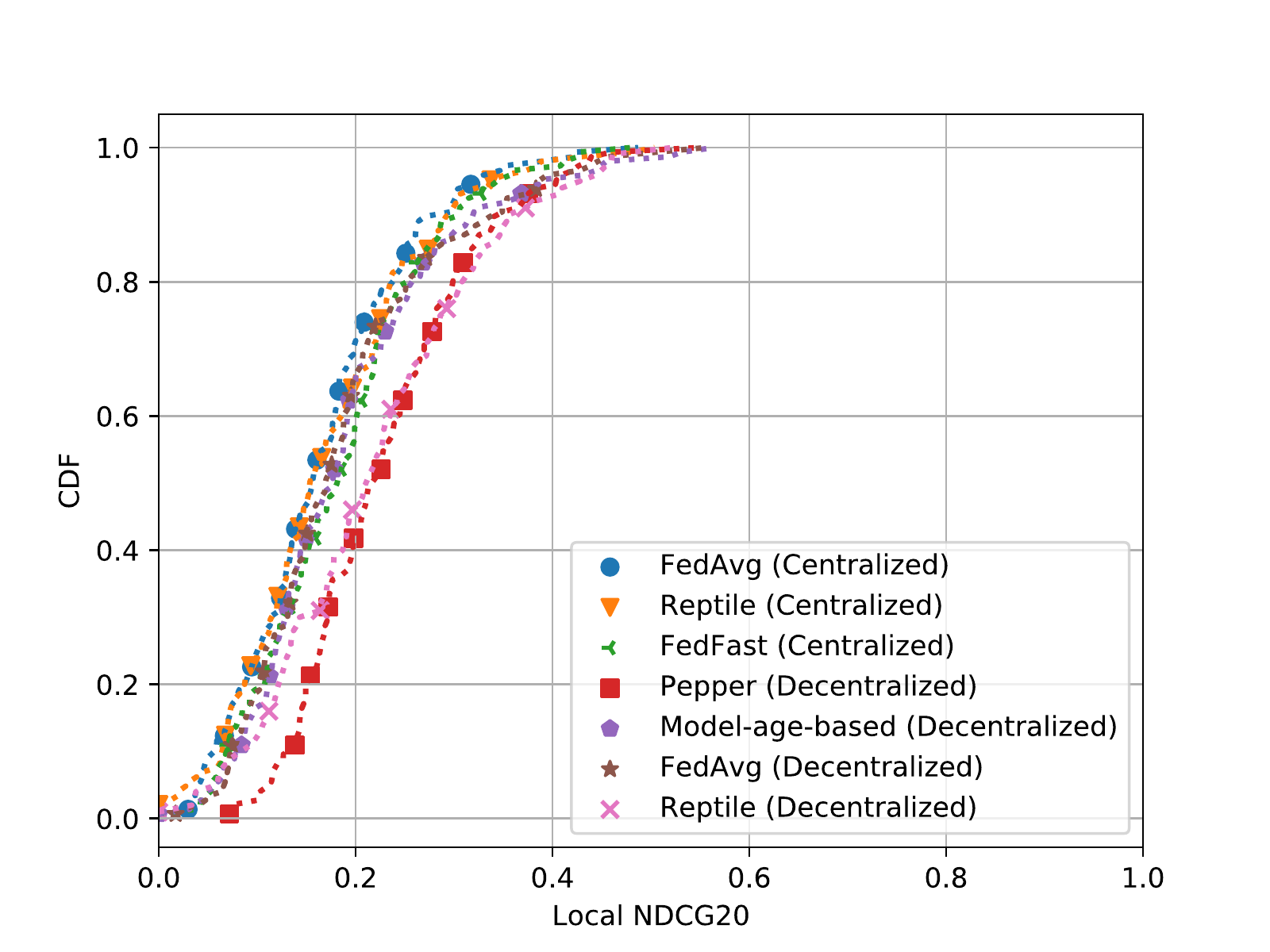}
    \caption{{Local NDCG@20 cumulative distribution function.}}
        \label{fig:percentilendcgfoursquare}
\end{subfigure}
    \caption{{Top-K recommendation quality distribution comparison on Foursquare-NYC (GMF, K = 20).}}
    \label{fig:percentilefoursquare}
\end{figure}

\begin{figure}[!htb]
\centering
\begin{subfigure}[b]{0.45\textwidth}
    \centering
    \includegraphics[scale=0.5]{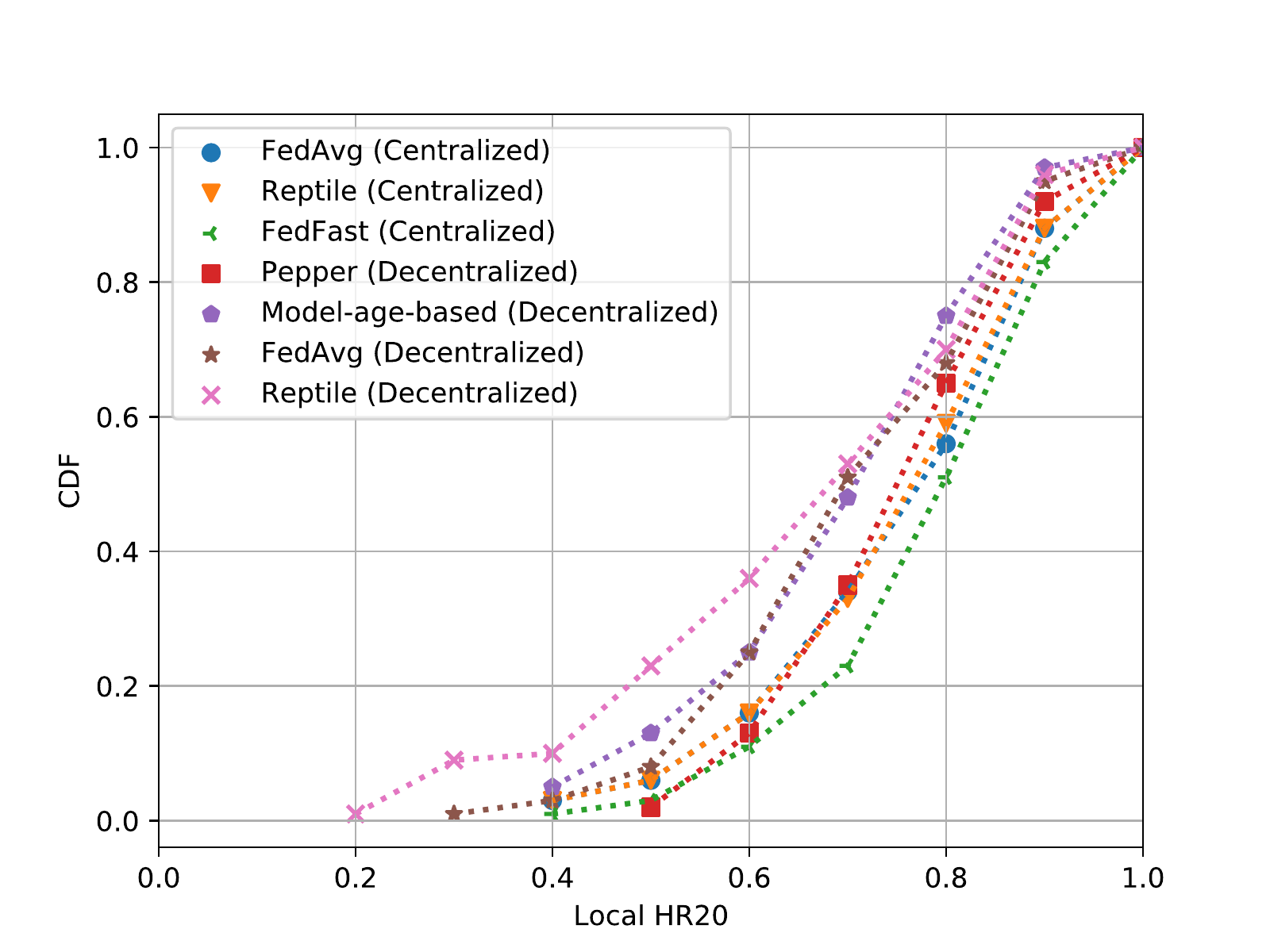}
    \caption{Local Hit Ratio@20 cumulative distribution function.}
    \label{fig:percentilehrmovielens}
\end{subfigure}
\hfill
\begin{subfigure}[b]{0.45\textwidth}
    \centering
    \includegraphics[scale=0.5]{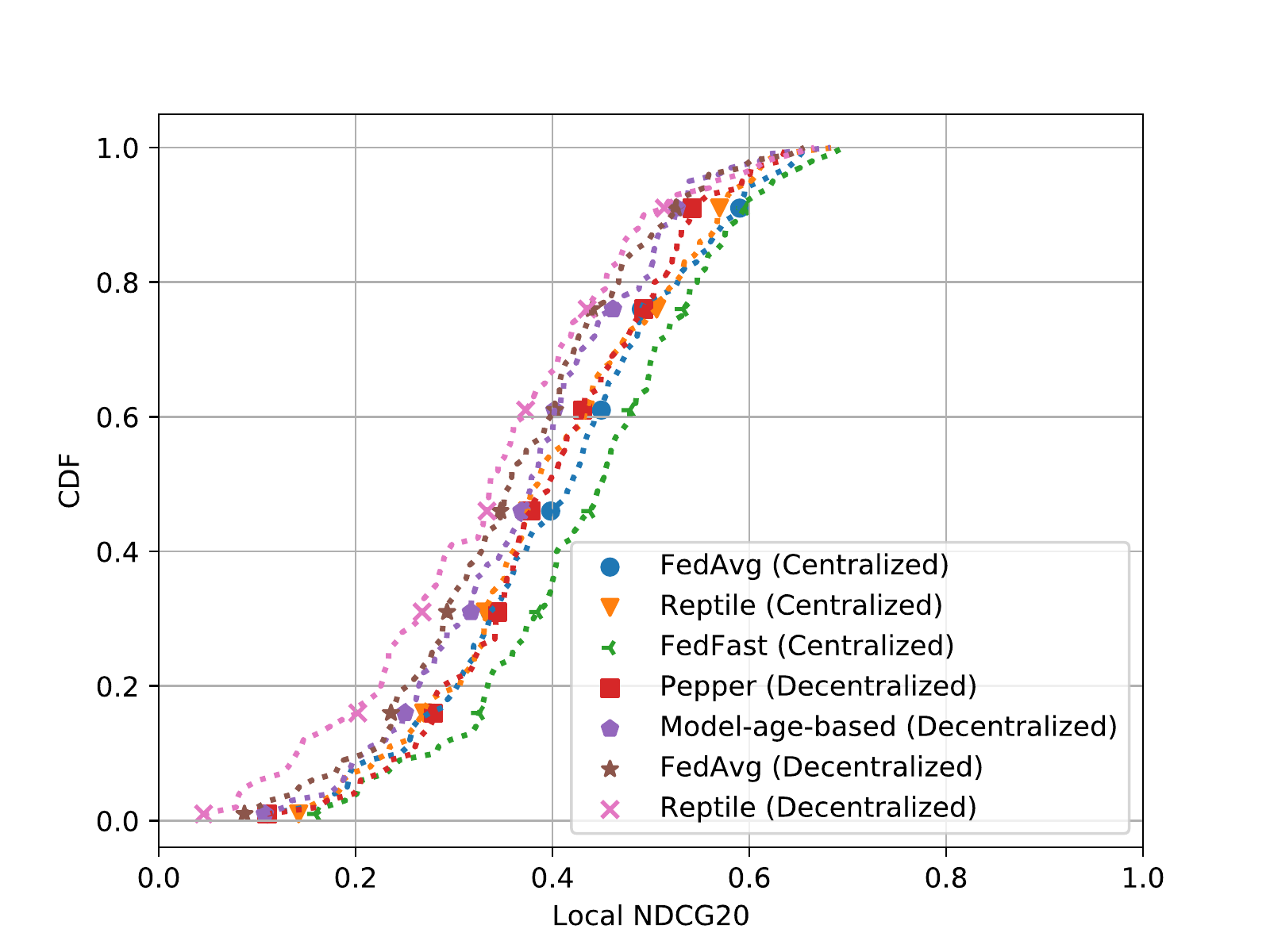}
    \caption{Local NDCG@20 cumulative distribution function.}
    \label{fig:percentilendcgmovielens}
\end{subfigure}
    \caption{{Top-K recommendation quality distribution comparison on MovieLens\hyp{100K} (GMF, K = 20).}}
    \label{fig:percentileML}
\end{figure}

In this experiment, we are interested in evaluating the individual satisfaction of each user. Figures~\ref{fig:percentilehrfoursquare} and~\ref{fig:percentilehrmovielens} illustrate the cumulative distribution functions of the HR in two datasets: Foursquare-NYC and MovieLens.

\begin{figure}[!htb]
\centering
\begin{subfigure}[b]{0.45\textwidth}
\centering
    \includegraphics[scale=0.5]{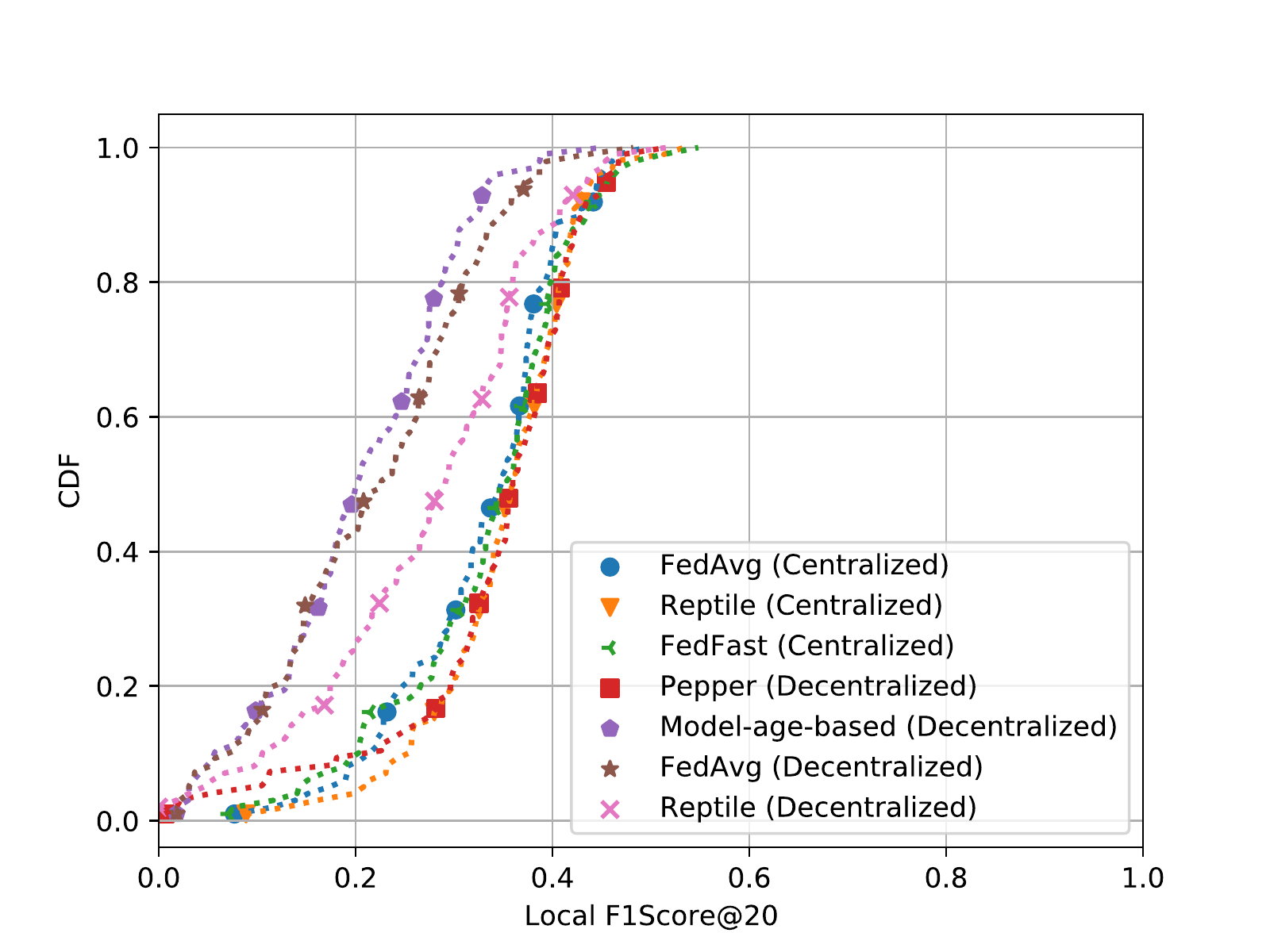}
    \caption{{Foursquare\hyp{NYC}.}}
    \label{fig:percentilef1scorefoursquare}
\end{subfigure}
\hfill
\begin{subfigure}[b]{0.45\textwidth}
\centering
    \includegraphics[scale=0.5]{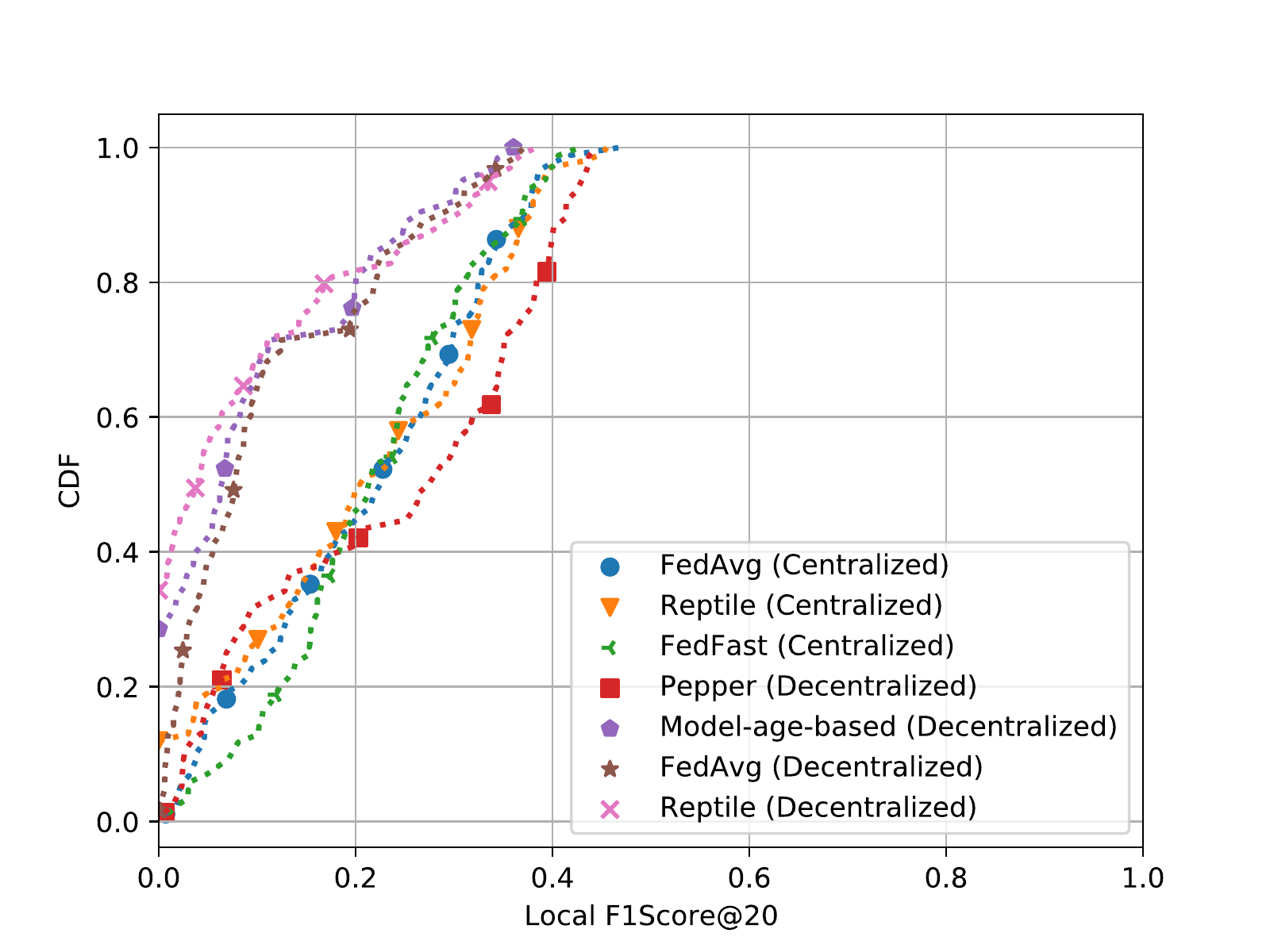}
    \caption{{Gowalla\hyp{NYC}.}}
    \label{fig:percentilef1scoregowalla}
\end{subfigure}
 \caption{{Top-K F1-Score cumulative distribution function (PRME-G, K = 20).}}
    \label{fig:percentileprmeg}
\end{figure}

Similarly, Figures~\ref{fig:percentilendcgfoursquare} and~\ref{fig:percentilendcgmovielens} illustrate that for NDCG. In these figures a point $(x=HR,y=CDF(x))$, respectively $(x=NDCG, y=CDF(x))$, represents the proportion of users $y$ having a hit ratio at most equal to HR, respectively an NDCG value at most equal to NDCG. Hence for a fixed value of HR or NDCG, the best curve is the one furthest to the right.
Centralized and generically aggregated models are often optimized with respect to a global objective and are, thus, less tailored to each individual user.
Therefore, there is often a large gap between the tail and the average performance of these solutions. \sysname aims to reduce this gap. For example, the 99.9th percentile in \sysname achieves about three times the performance of its best competitor~(\ie, centralized FedAvg) on Foursquare (see Figure~\ref{fig:percentilehrfoursquare}).
We mention that the 99th percentile represents the values corresponding to a CDF of 0.1 on Figures~\ref{fig:percentilefoursquare},~\ref{fig:percentileML}, and~\ref{fig:percentileprmeg}.
Concerning MovieLens, \sysname is generally overpassed by centralized solutions on both short and long tail performance. However, it always achieves better tail performance than its decentralized competitors (see Figure~\ref{fig:percentileML}). The aforementioned experiments are performed with a Generalized Matrix Factorization (GMF) model. Figures~\ref{fig:percentilef1scorefoursquare} and~\ref{fig:percentilef1scoregowalla} illustrate the percentiles' F1 score results for the PRME-G model. Similarly to the previous use case, The FedFast protocol and Reptile have the best short tail score. Nevertheless, we can observe that \sysname has a performance close to the best performing solution and even outperforms them on the long tail (\ie, 60th to 85th percentile). For the short tail, around ~2.5\% of nodes don't seem to have enough datapoints to converge PRME-G in a decentralized learning setup so the 99.9th percentile F1 score of all decentralized solutions tends towards zero.  
In summary, the evaluation results show that  \textbf{\sysname outperforms state\hyp{of}\hyp{the}\hyp{art} decentralized solutions on long and short tail performance}. Moreover, \sysname slightly improves the short tail (\ie, 99.9th percentile) on MovieLens over centralized competitors, while significantly improving the long tail on the other datasets~\textbf{(RQ2)}.

\subsubsection{{Effects of Sparsity}}
\label{subsec:sparsity}
Sparsity is a very important aspect that often negatively impacts the performance of recommendation systems.
Therefore, to evaluate its impact on \sysname, we have generated two additional datasets (Dense ML-100k and Sparse ML-100k) by clustering the users of the original MovieLens-100k using k-means. Dense ML-100k is composed of the users which come from the most dense cluster while Sparse ML-100k is composed of all of the outliers.

\begin{figure}[!h]
\centering
    \includegraphics[width=0.6\textwidth]{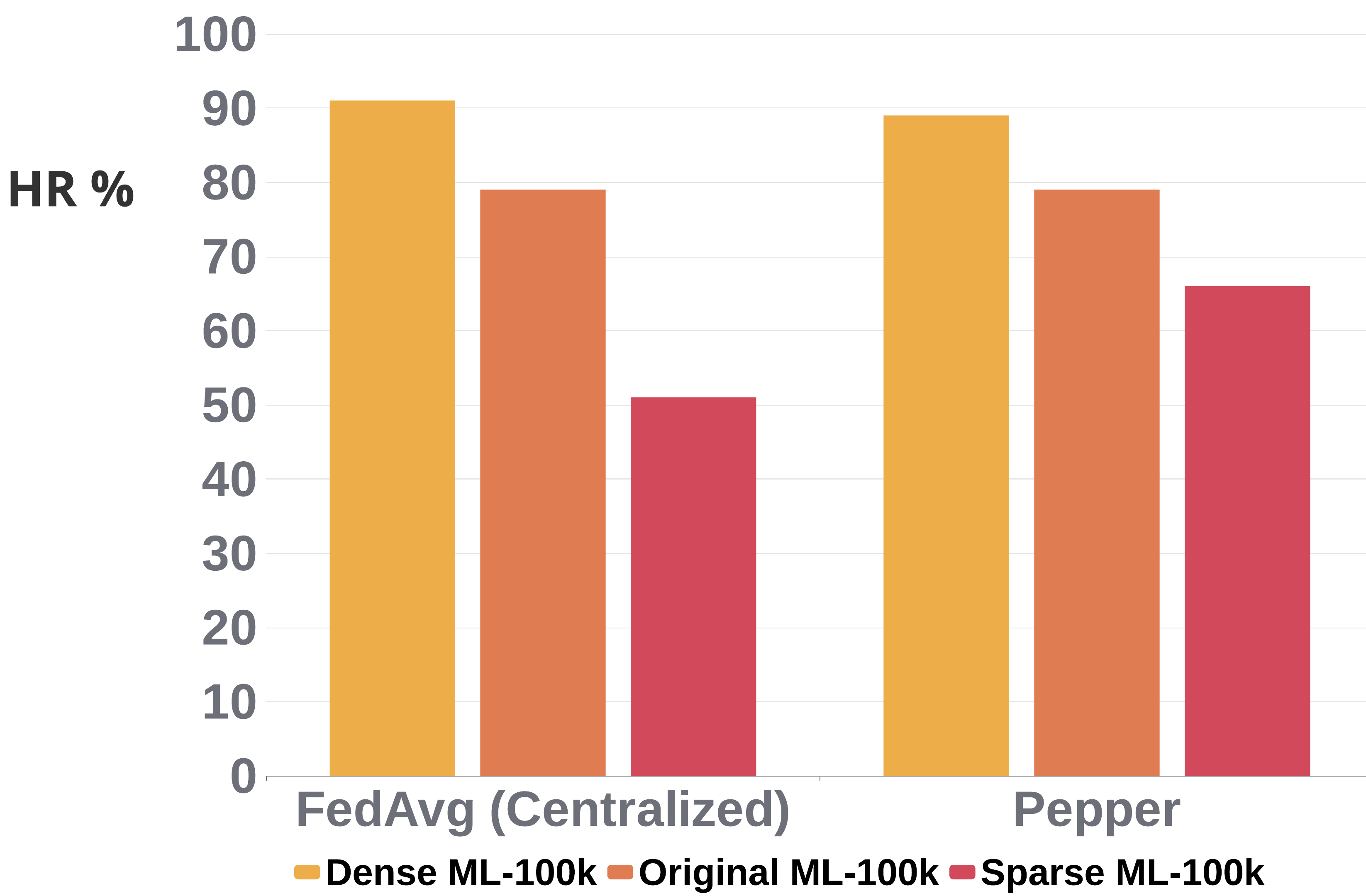}
\caption{{The robustness of \sysname to sparsity in comparison with Federated Averaging, on three different sparsity levels: Dense ML-100k (sparsity = 0.64, users = 579), Original ML-100k (sparsity = 0.93, users = 943) and Sparse ML-100k (sparsity = 0.99, users = 201).}
    }
    \label{fig:sparsityfig}
\end{figure}

Higher sparsity levels can often lead to users having less overlap in terms of preferences (\ie, rated movies, visited locations), which makes it challenging for global models, and especially those based on learning the user-item relationships, to fit all users' preferences. Therefore, the centralized solutions (\eg, FedAvg), which train a unique model will experience more difficulties to provide satisfactory results for all users. However, as shown in Figure~\ref{fig:sparsityfig}, \textbf{the personalized models built by \sysname, through their ability to capture the preferences of each user, seem to be more robust to high degrees of sparsity}. This result further corroborates the  Section~\ref{subsec:averageperf} results, where \sysname was more competitive with the centralized approaches on sparsest datasets~\textbf{(RQ3)}.

\subsubsection{{Sensitivity of \sysname to the exploitation\hyp{exploration} ratio alpha}}
\label{subsec:alphaparameter}
\begin{figure}[!htb]
\centering
\begin{subfigure}[b]{0.45\textwidth}
  \centering
    \includegraphics[scale=0.5]{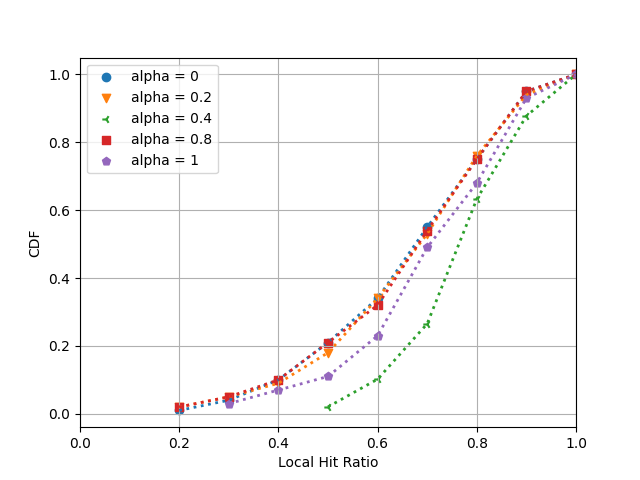}
    \caption{{Cumulative Distribution Function comparison between different alpha values on MovieLens\hyp{100k} (GMF).}}
    
    \label{fig:averagealpha}
\end{subfigure}
\hfill
\begin{subfigure}[b]{0.45\textwidth}
   \centering
    \includegraphics[scale=0.28,width=\textwidth]{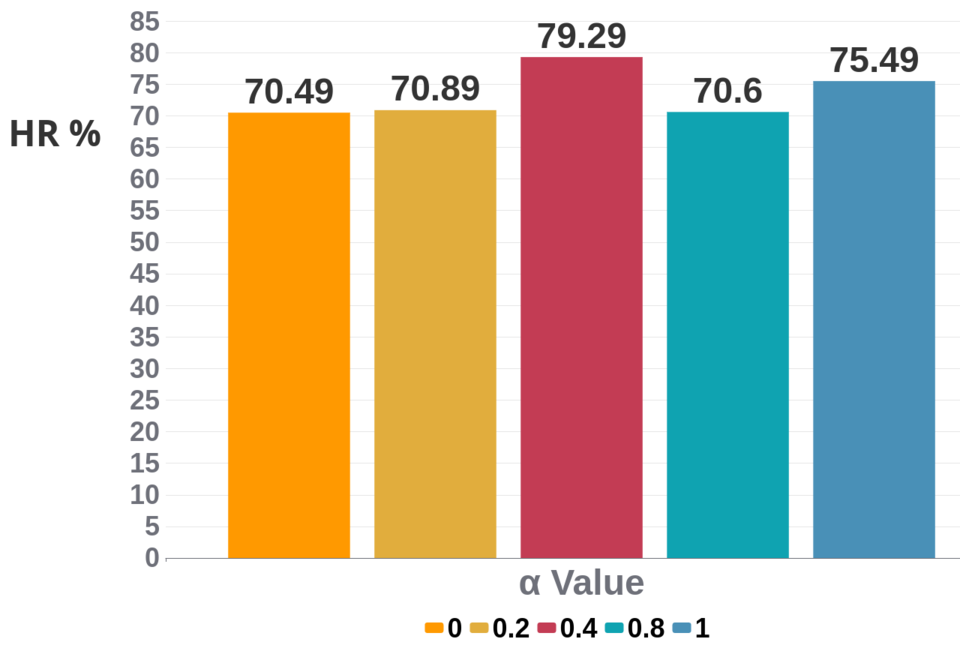}
    \caption{{Average performance w.r.t different values of Alpha on MovieLens\hyp{100k} (GMF).}}
    \label{fig:percentilealpha}
    \end{subfigure}
    \caption{{Sensitivity of \sysname to the exploitation-exploration ratio alpha.}}
    \label{fig:alpha}
\end{figure}

{Peer\hyp{sampling} is a key component of the system as the search for similar neighbors helps model personalization and hence improves the local performance. In this context, the alpha parameter in \sysname, which fixes the exploration versus the exploitation ratio, plays a crucial role. We have evaluated the impact of this parameter on the performance of \sysname in an equal number of training rounds by measuring the CDF of the local hit ratio with various values of alpha, as well as computing the average hit ratio. Results depicted in Figure~\ref{fig:averagealpha} show that a value of alpha = 0.4 yields the best average hit ratio. By looking closer at Figure~\ref{fig:percentilealpha}, we observe that for a value of alpha = 0 where nodes never change their neighbors, there are two consequences: i) the rate of model dissemination is reduced and ii) the initial placement of nodes becomes very significant, as a node without similar first or second degree neighbors will not be able to personalize its model and will have to settle for the models it receives, finding itself forced to build a more global model. On the other hand, for an alpha = 1, where at each peer-sampling period, nodes completely change their view, which does not give them the time to make the most of similar nodes' model, it seems that personalization is significantly diminished but not completely annihilated. This leads to better models than the first case. Inevitably, the average performance suffers in both of these extreme cases (70.49\% in the former and 75.49\% in the latter). This also shows, as expected, that an aggregation maximizing local performance cannot have a significant impact if the model evaluation step is not exploited to identify similar nodes and take advantage of their models. Simultaneously, \sysname cannot entirely rely on a personalized peer\hyp{sampling} as randomization is important to disseminate models and to find better neighbours~\textbf{(RQ4)}.}

\subsubsection{{Sensitivity of \sysname to the peer set size}}
{In order to assess the impact of the peer set size on the performance of \sysname, we performed an experiment where we doubled the peer set size used in the other experiments of this paper (\ie, peer set size = 6 instead of peer set size = 3). We performed this experiment on the MovieLens-100k dataset using the GMF model. Results are depicted in Table~\ref{tab:peersetsize}. From these results, we observe that increasing the peer set size positively impacts  convergence as the models need less rounds to converge. Moreover, performance is faintly improved. However, this comes at the price of an increased computational overhead as nodes receive more models, hence, computing more SGD and aggregation steps. To choose our peer set size, we complied to the theoretical bounds proved by previous works on gossip~\citep{Bollobs2004TheDO}. The latter state that a logarithmic peer set size in the total number of nodes provides the best tradeoff between all of these metrics under the condition that each {new node $v$ is connected to another one $w$ with a probability proportional to the number of neighbours of $w$~\textbf{(RQ5)}.}}

\begin{table}[!htb]
\centering
\resizebox{0.9\textwidth}{!}{
   \begin{tabular}{|| c|c|c|c ||}
         \hline
         & Average Performance (HR \%) & Average Communication Rounds & Average Overhead (seconds)\\
        \hline\hline
        Peer Set Size = 3 & 0.79 & 283 & 53.4\\
        \hline\hline 
        Peer Set Size = 6 & 0.80 & 136 & 119.8\\
         \hline
    \end{tabular}
}    

\caption{{Impact of the Peer Set Size on \sysname.}}
\label{tab:peersetsize}
\end{table}

\subsubsection{Overhead evaluation}
\begin{figure}[!h]
\centering
\begin{subfigure}[b]{0.45\textwidth}
    \centering
    \includegraphics[scale=0.5]{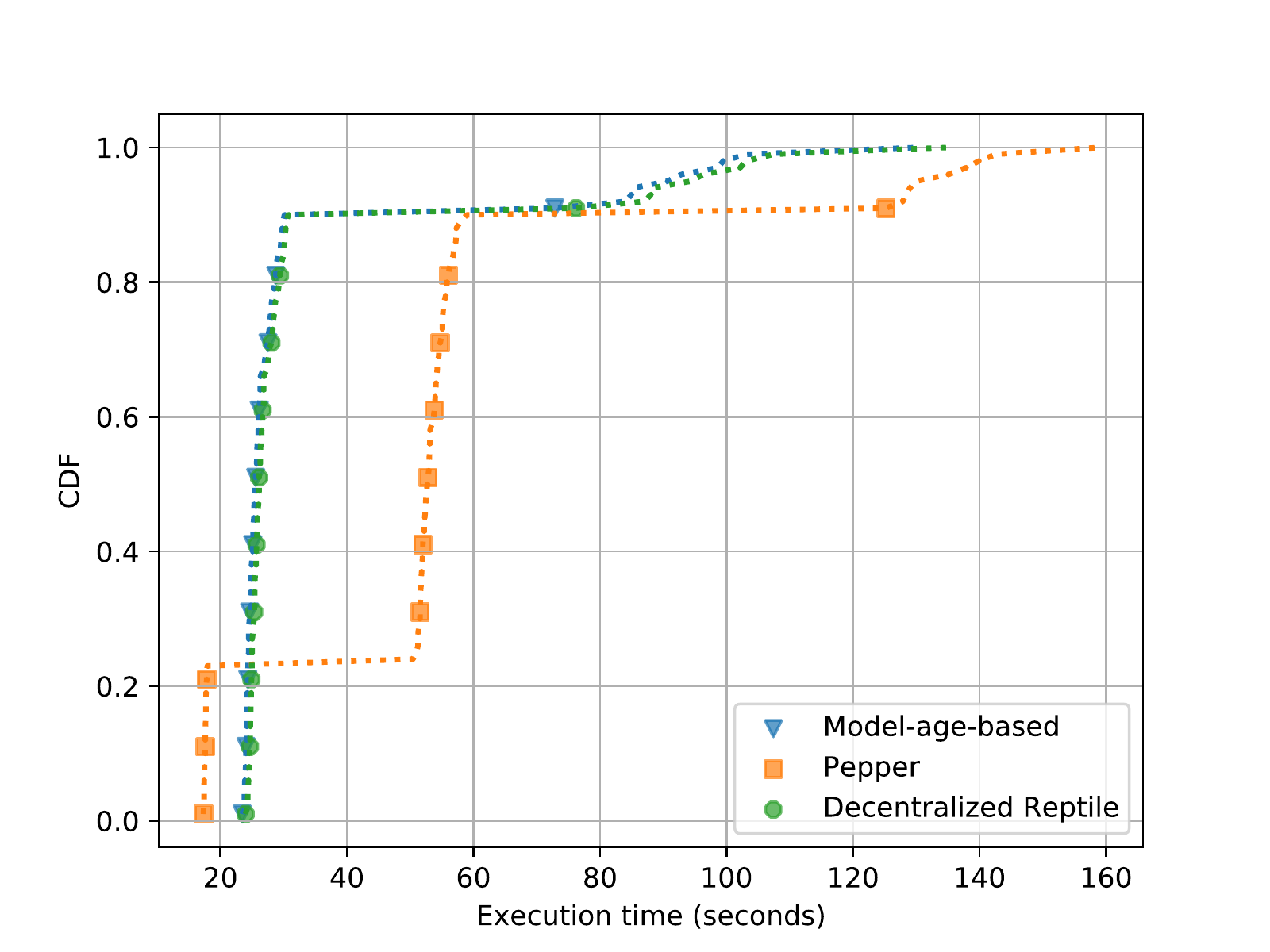}
    \caption{The CDF of the total computing time for each individual node until the model convergence is reached. }
    \label{fig:overheadfig}
\end{subfigure}
\hfill
\begin{subfigure}[b]{0.45\textwidth}
    \centering
    \includegraphics[scale=0.5]{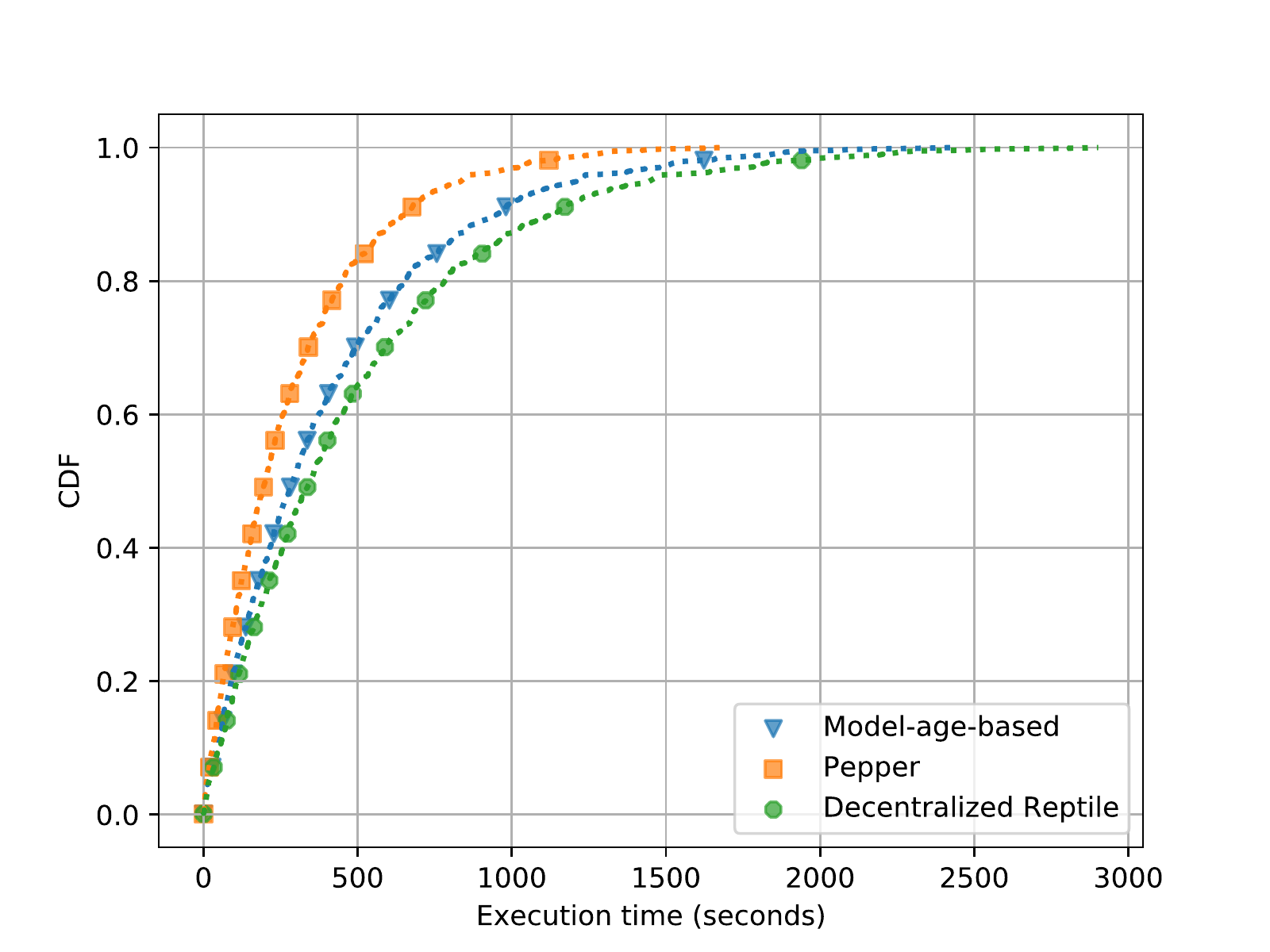}
    \caption{The average number of communication rounds executed by each individual user.}
    \label{fig:communicationrounds}
    \end{subfigure}
\vfill
\caption{Computation vs Communication costs.}
\end{figure}

We also evaluated the overhead of \sysname compared with its decentralized competitors.
In Figure~\ref{fig:overheadfig} we represented the CDF of the total computation time for each individual node. 
On each node, we sum up the time spent to compute the SGD algorithm on local data and the aggregation time over all of the learning process and then, we represent these values as a CDF. In this latter, we did not consider the Decentralized FedAvg algorithm due to it having the same cost as Model\hyp{age}\hyp{based}.
We can observe that ~90\% of the nodes in both Decentralized Reptile and Model-age-based have fairly homogeneous execution times, with ~90\% of them having around ~25 seconds.
\textbf{Concerning \sysname, ~21\% of the nodes have a total execution time inferior to the other decentralized solutions and ~69\% of the nodes have a longer execution time of ~54 seconds (RQ6).}
These two clusters of users are formed because of the following phenomenon.
The nodes with less data will usually send less-accurate models to their neighbours so, because of our Personalized Peer-sampling, they will be chosen with less probability by the other nodes. 
Therefore, they will receive less models to be evaluated by \sysname and therefore their execution time will be shorter.

Even if the execution time is higher on some nodes, \sysname personalizes the models to the preferences of each user so it needs less communication rounds to converge (see Figure~\ref{fig:communicationrounds}). 
\textbf{On average, the nodes of \sysname reach the model convergence after 283 communication rounds comparing with 410 for Model-age-based and 490 for Decentralized Reptile}. When analysing the high percentiles, we can observe that the gap increases even more. For example, the 99th percentile of \sysname is characterized by 799 and 1247 fewer rounds than Model-age-based and Decentralized Reptile, respectively~\textbf{(RQ6')}.

\section{Discussion}
\label{sec:discussion}
In this section we discuss the limitations of \sysname and their possible mitigations.\\

\emph{On the use of simulations to evaluate \sysname}: 
Decentralized systems imply the existence of a large number of nodes interconnected in a peer-to-peer network.
However, few organizations have access to such a system in the physical world. 
Therefore, many research works rely on simulators to evaluate their decentralized algorithms~\cite{DecentralizedRecommendationBasedonMatrixFactorization:AComparisonofGossipandFederatedLearning,Decentralizedlearningworks:Anempiricalcomparisonofgossiplearningandfederatedlearning}. 
In \sysname, we rely on Omnet++~\cite{omnetpp}, a popular network simulator allowing to evaluate settings involving up to 1000 nodes. Practical studies have further assessed the accuracy of using this simulator compared to real testbeds~\cite{colesanti2007accuracy}.\\

\emph{On the privacy of \sysname}:
We assume in this paper that participants are trusted. However, despite the fact that nodes keep their data in their premises, there exist attacks that have been devised in the context of Federated Learning and that can leak private information from the exchanged model updates~\cite{gong2016you,nasr2019comprehensive}. These attacks could very well take place in a fully decentralized setting as in \sysname, but considering such attacks is out of the scope of this paper. Nevertheless, there exist research works focusing on this issue (e.g., solutions relying on differential privacy~\cite{naseri2020local}). Specifically, \citet{hegedHus2017differentially} have investigated the use of differential privacy in the context of Gossip Learning~\cite{hegedHus2017differentially}. In the more generic context of Federated Learning other solutions to protect users against inference attacks are heavily investigated by the research community as surveyed in~\cite{enthoven2021overview}. A combination of existing solutions with \sysname shall be investigated in future work.\\

\emph{On the resilience of \sysname to poisoning attacks}: Similarly to inference attacks, there exist poisoning attacks that have been proposed in the context of Federated Learning and that could easily be launched by \sysname participants (\eg, \cite{huang2021data,tolpegin2020data,yin2018byzantine}). For instance, malicious participants could send poisonous model updates to their neighbors. While we considered this issue as being out of the scope of this paper, the fact that nodes in \sysname locally assess the relevance of received models before aggregating them could naturally protect them against such attacks. We plan to assess in our future work the resilience of \sysname to poisoning attacks. In case these attacks are still possible, there exist model aggregation functions that have been devised in the context of FL in order to identify and exclude poisonous gradients and that could be integrated in \sysname (\eg, \cite{tianxiang2019aggregation,blanchard2017machine, yin2018byzantine}).\\

\emph{On the impact of churn, network dynamics and node heterogeneity}: Churn and more generally network dynamics and node heterogeneity (\eg, in terms of computing and networking capabilities) have an impact on the performance of gossip protocols and thus on the applications running on top of them such as \sysname. While studying the impact of churn, network dynamics and node heterogeneity is important, we considered it out of the scope of this paper. In practice, we expect churn, network dynamics and node heterogeneity to have a similar impact on \sysname and on its decentralized competitors, who did not either look at these issues yet. Nevertheless, the distributed systems community has investigated many gossiping flavors that increase the robustness of gossip to the above issues. For instance,~\citet{frey2009heterogeneous}
showed that while the gossip protocol is by design robust to churn in a uniform and highly capable distribution, it can be adapted to heterogeneous setups by adjusting the view size of each node proportionally to its upload capability. Combining this approach with theoretical results, which show that an average logarithmic view size preserves the dissemination of a gossip protocol,~\cite{frey2009heterogeneous} enhances considerably the robustness of gossip protocols to churn in heterogeneous contexts. More recently, preliminary solutions have been considered for gossip learning (\eg,~\cite{han2020accelerating,giaretta2019gossip}). For instance, in~\cite{giaretta2019gossip}, a simple yet effective mechanism that skips the aggregation and update phases when the view size of the sending node is small, ensures that models coming from low-degree nodes propagate faster. As opposed to this approach, \citet{han2020accelerating} proposed a data-driven one where the ratio and the nature of training data taken in consideration by a node is adjusted accordingly to its computing and networking abilities. A combination of such solutions with \sysname shall be investigated in future work.

\section{Related Work}
\label{sec:rel_works}

In the past several solutions have been proposed to build distributed recommender systems. The first ones focused mainly on memory-based methods of recommendations, which usually use similarity metrics to build communities of users and/or items.  In addition to being highly time consuming~\cite{recsyssurvey}, these methods require sharing recommendation information (\eg, ratings, user-items interactions, etc.) between users which was done either through distributed hash tables~\cite{AscalableP2Precommendersystembasedondistributedcollaborativefiltering,Auserorientedcontentsrecommendationsysteminpeertopeerarchitecture} or directly via gossip protocols~\cite{APeer-to-PeerRecommenderSystemwithPrivacyConstraints,Tribler}. This information sharing is sometimes transgressing user privacy so it motivated more careful designs such as~\cite{OnGossipbasedInformationDisseminationinPervasiveRecommenderSystems,hashem2018crowd}, which imply a notion of trust between subsets of users. However, these setups were neither practical nor generalizable so various privacy-preserving recommender systems started to emerge. 

The privacy-preserving recommender systems fall into three categories. 
First, there are \emph{homomorphic-encryption-based} solutions~\cite{wang2019,kim2016efficient,guerraoui2017}, where user profiles, ratings, and any sensitive information is encrypted before being processed. This approach is computationally expensive and introduces a significant overhead. A state-of-the-art implementation of fully homomorphic encryption~\cite{Rohloff2014} requires 28 hours to make a recommendation on MovieLens ~\cite{movielens}.
The second category is composed of \emph{solutions based on differential privacy (DP)}~\cite{gao2020,gao2019privacy,Privacy-preservingdistributedcollaborativefiltering,mcsherry2009} where user profiles are obfuscated with noise. However, due to noise interference, DP often introduces a compromise between the privacy and the accuracy of recommendations. 
The third category of privacy-preserving recommender systems is represented by the \emph{Federated Learning-based solutions}. Depending on the presence of parameter servers or not, Federated Recommender Systems can be divided in two categories:

\emph{Centralized FL recommender systems}. These systems are characterized by the presence of a centralized server which coordinates the learning process. In the works of \citet{ANovelPrivacyPreservedRecommenderSystemFrameworkbasedonFederatedLearning} this entity allows the users to collaboratively train a recommender model based on sensitive information such as location and age which are kept private on client devices. In addition, user-items interactions are considered to be publicly available so they are centralized and leveraged to train another model on the server-side. However, it can be argued that such interactions are also private by nature. Therefore, other works~\cite{Fast-adaptingandPrivacy-preservingFederatedRecommender,ASimpleandEfficientFederatedRecommenderSystem,muhammad2020fedfast} are more strict in this regard as they do not assume any publicly available data. 
For personalization and user satisfaction purposes, two distinct approaches can be identified : Meta\hyp{Learning}\hyp{based} solutions and similarity\hyp{based} ones. The former~\cite{Fast-adaptingandPrivacy-preservingFederatedRecommender,ASimpleandEfficientFederatedRecommenderSystem,fallah2020personalized} are based on the idea of training a global model following a traditional method (\eg, Federated Averaging) then fine\hyp{tuning} it on local data by making meta\hyp{steps}, which turn the global model into a more personalized one. While such techniques have been found to be effective in many use cases, they incur a risk of overfitting due to the purely local personalization. In contrast, in \sysname, a model is personalized by taking advantage of similar models, which reduces the risk of overfitting. On the other hand, these Meta\hyp{Learning} techniques depend considerably on having a good global model as a starting point. For that reason, they have not, to the best of our knowledge, been considered in decentralized contexts, where obtaining such a global model is often fairly difficult. In the similarity\hyp{based} class of solutions, FedFast~\cite{muhammad2020fedfast} can be considered as one of the most influential works. It incorporates a similarity-based client sampling technique employed to cluster users so that an active aggregation function can be applied within these clusters, which improves the performance and convergence speed. However, the client sampling technique is based on a per\hyp{round} clustering, which, as stated by the authors, is quite costly (\eg, a time complexity $\mathcal{O}(n^{81})$ per round on MovieLens\hyp{100k}). In contrast, \sysname emulates a cheaper and gradual clustering along the learning process, by keeping similar nodes close to each other. Another difference is that \sysname considers all models and only ponders their magnitude with respect to their similarity, while in contrast, FedFast completely excludes parts of the models that are outside of the cluster of a user. 
Finally, FedFast also inherits the main drawback of centralized FL solutions which is the presence of a central server that can face different kinds of challenges (See Section \ref{subsec:fl}).

\emph{Decentralized FL recommender systems.} In decentralized FL recommender systems, there is no central server and clients instead exchange models with each other. To the best of our knowledge, the work of \citet{DecentralizedRecommendationBasedonMatrixFactorization:AComparisonofGossipandFederatedLearning} is the only decentralized Federated Learning recommender system solution in the literature. In this work, the authors show that decentralization's impact on performance can be mitigated. For that purpose, different model compression methods as well as algorithmic enhancements in the form of flow control mechanisms were presented. While they do not specifically target aggregation functions, they still implement an aggregation function that takes into consideration the age of models, which allows models that have seen more data to have a larger weight. We consider this work as a baseline in our current evaluation settings to quantify the
added value of our performance-based aggregation function.

\section{Conclusion}
\label{sec:conclusion}
 In this work, we presented \sysname, a decentralized and privacy\hyp{preserving} recommender system. \sysname relies on Gossip Learning principles for enabling nodes to asynchronously train a model that better responds to their needs. At the heart of \sysname resides two key components: i) a personalized peer\hyp{sampling} protocol, which allows each node to keep in his neighborhood a proportion of similar nodes (taste wise), leading to a gradual clustering of users and ii) a simple yet effective model aggregation function that builds a model that is better suited to each user. We implemented \sysname on a networking simulator and evaluated its performance using three real datasets involving up to 1000 nodes and implementing two use cases: a location check-in recommendation and a movie recommendation. Our results show that, on average, nodes in \sysname converge with up to 42$\%$ less communications rounds than with other decentralized approaches while providing up to 8$\%$ improvement on average performance and up to 30$\%$ improvement on tail performance compared to decentralized competitors. As part of our future work we plan to investigate the impact of fully decentralizing recommender systems as in \sysname on their resilience to adversaries.

\bibliography{main}
\bibliographystyle{ACM-Reference-Format}
\appendix
\section{Individual Performance Comparison}
\label{appendix:others}

\begin{figure}[!htb]
\centering
\begin{subfigure}{0.45\textwidth}
    \centering
    \includegraphics[width=\textwidth]{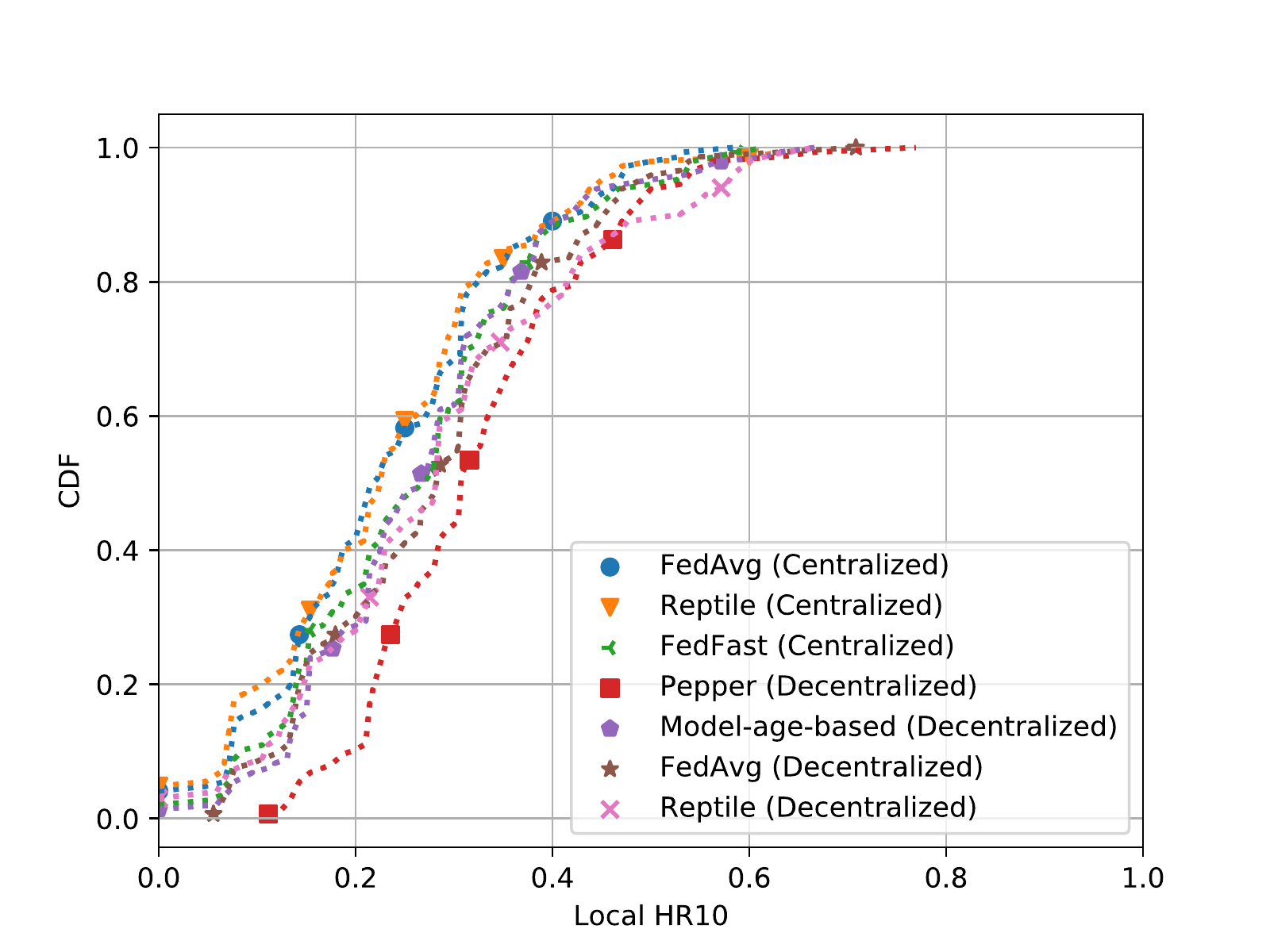}
    \caption{Local Hit Ratio@10 cumulative distribution function.}
    \label{fig:percentilehrfoursquare10}
\end{subfigure}
\hfill
\begin{subfigure}[]{0.45\textwidth}
    \centering
    \includegraphics[width=\textwidth]{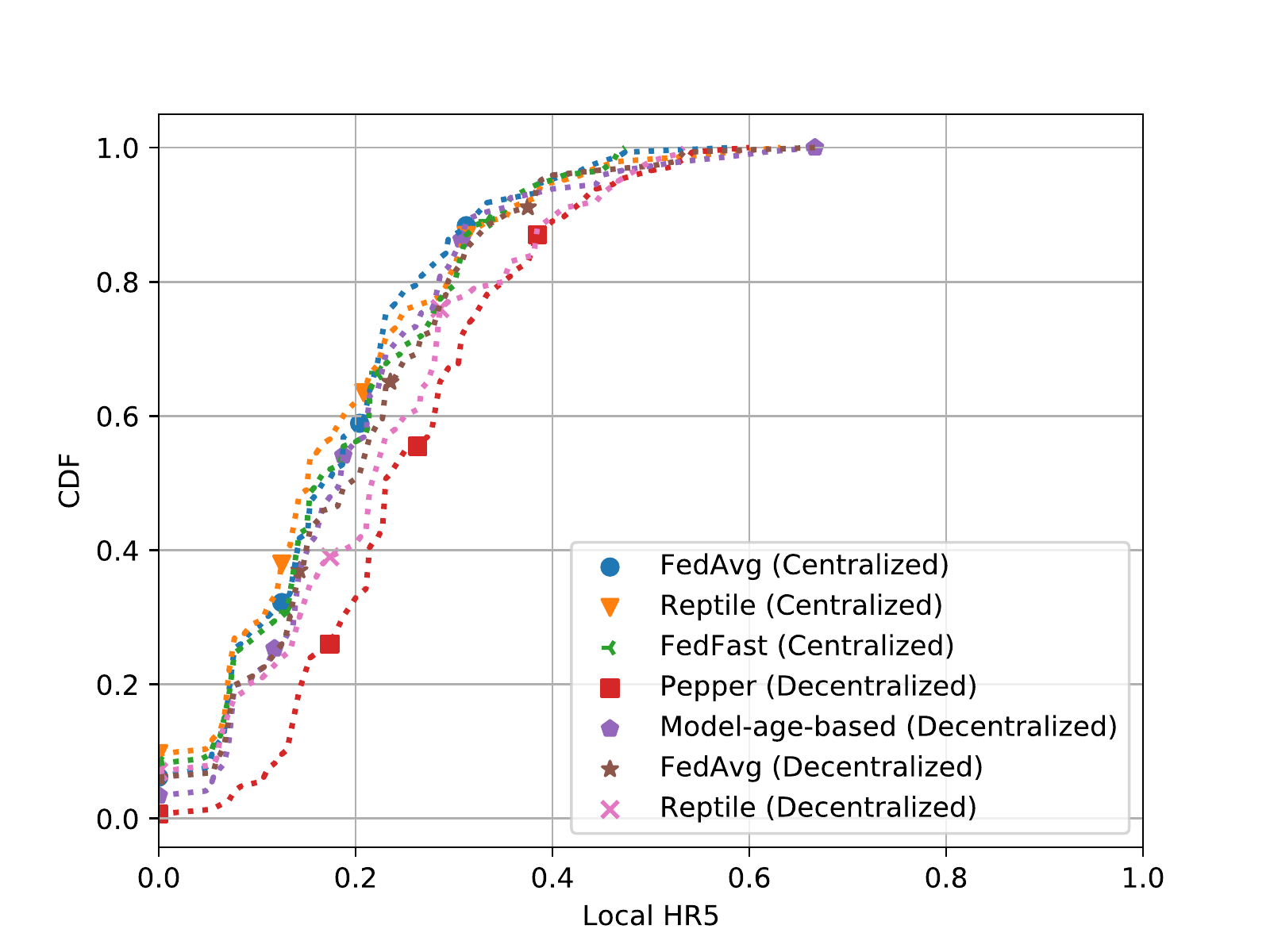}
    \caption{Local Hit Ratio@5 cumulative distribution function.}
    \label{fig:percentilehrfoursquare5}
\end{subfigure}
 \caption{ Top-K recommendation quality distribution comparison on Foursquare-NYC (GMF, K = 10 and K = 5).}
\end{figure}

\begin{figure}[!htb]
\centering
\begin{subfigure}[]{0.45\textwidth}
    \centering
    \includegraphics[width=\textwidth]{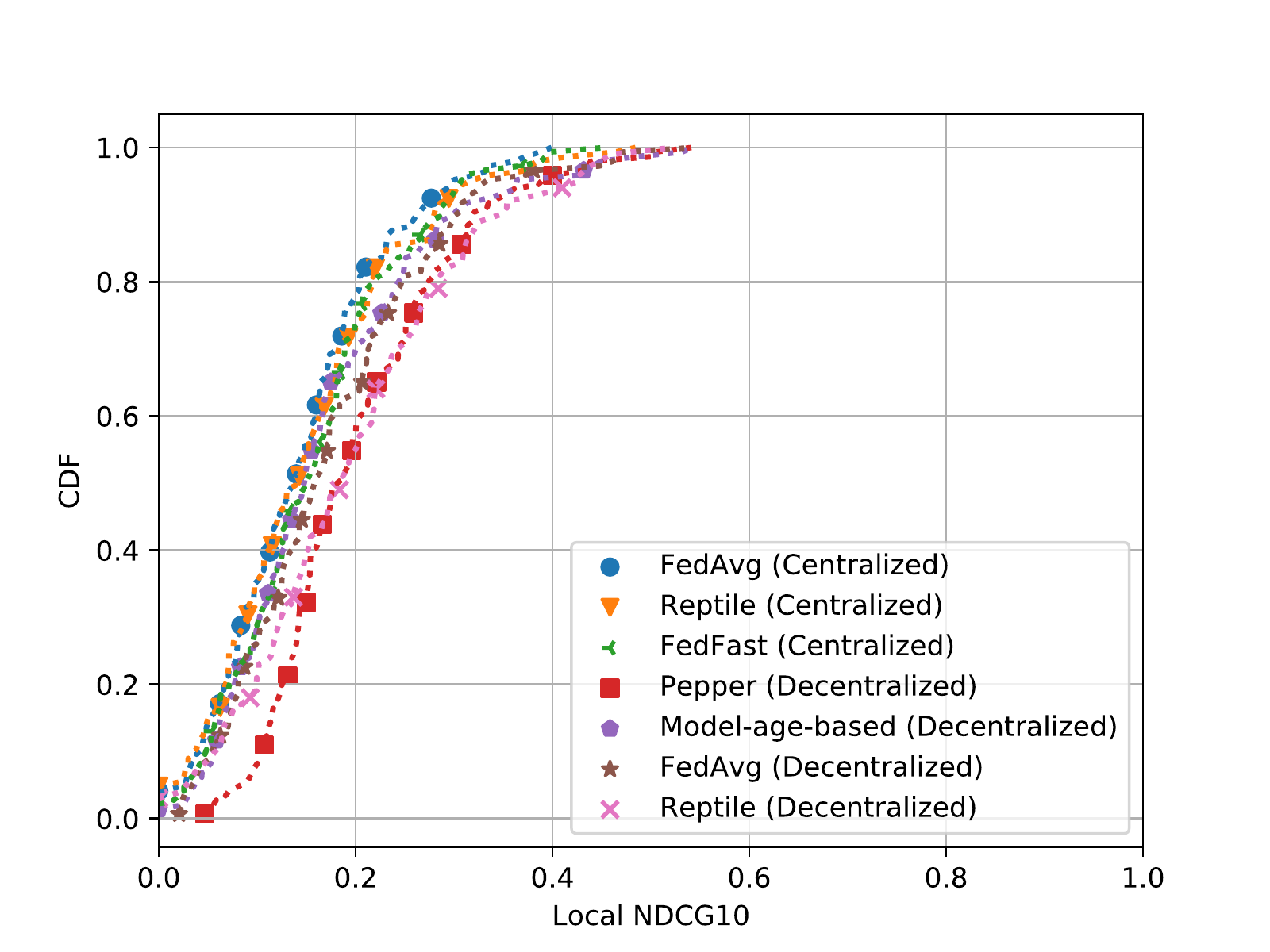}
    \caption{Local NDCG@10 cumulative distribution function.}
    \label{fig:percentilendcgfoursquare10}
\end{subfigure}
\hfill
\begin{subfigure}[]{0.45\textwidth}
    \centering
    \includegraphics[width=\textwidth]{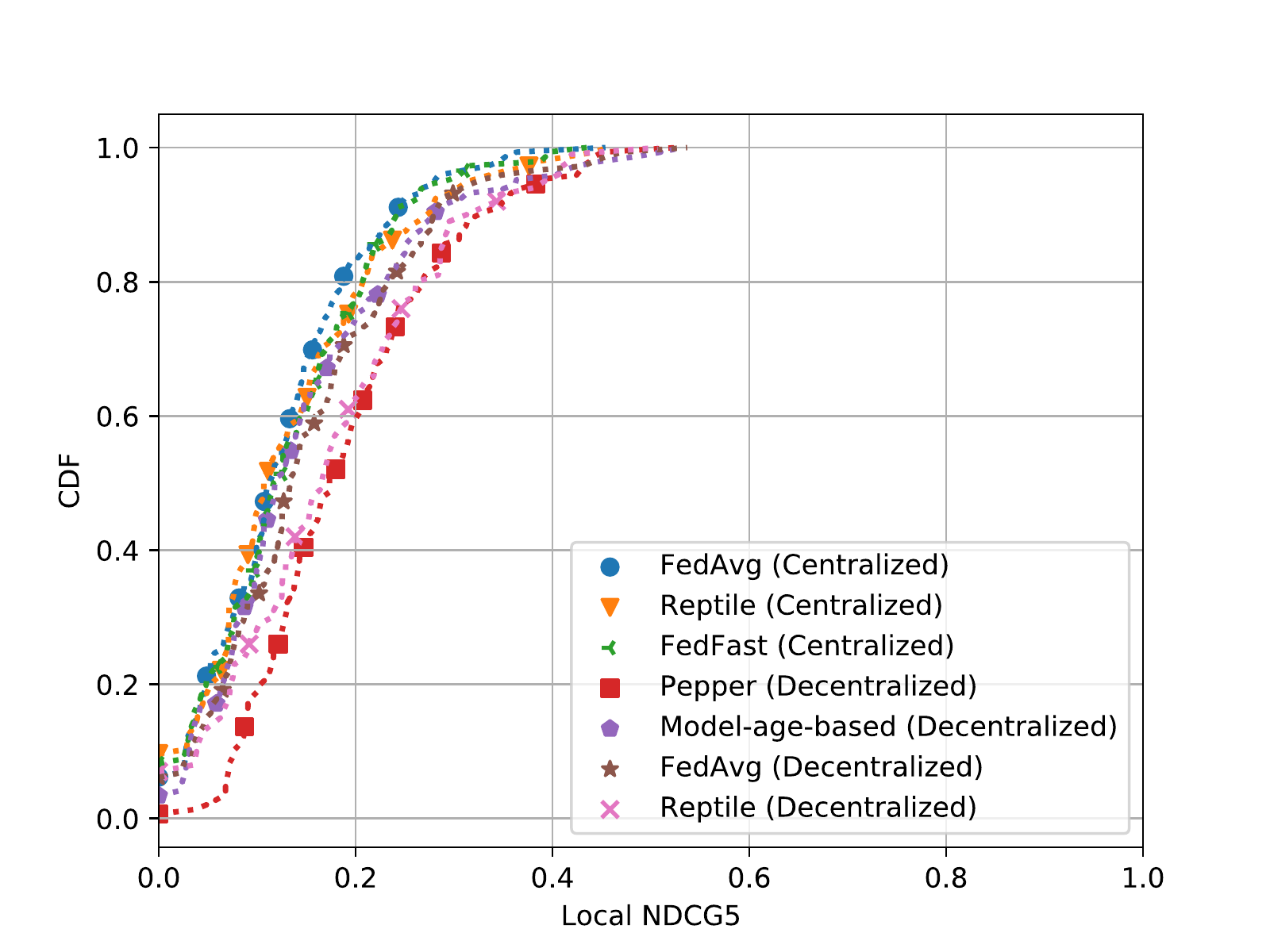}
    \caption{Local NDCG@5 cumulative distribution function.}
    \label{fig:percentilendcgfoursquare5}
\end{subfigure}
\caption{ Top-K recommendation quality distribution comparison on Foursquare-NYC (GMF, K = 10 and K = 5).}
\end{figure}

\begin{figure}[!htb]
\centering
\begin{subfigure}[]{0.45\textwidth}
    \centering
    \includegraphics[width=\textwidth]{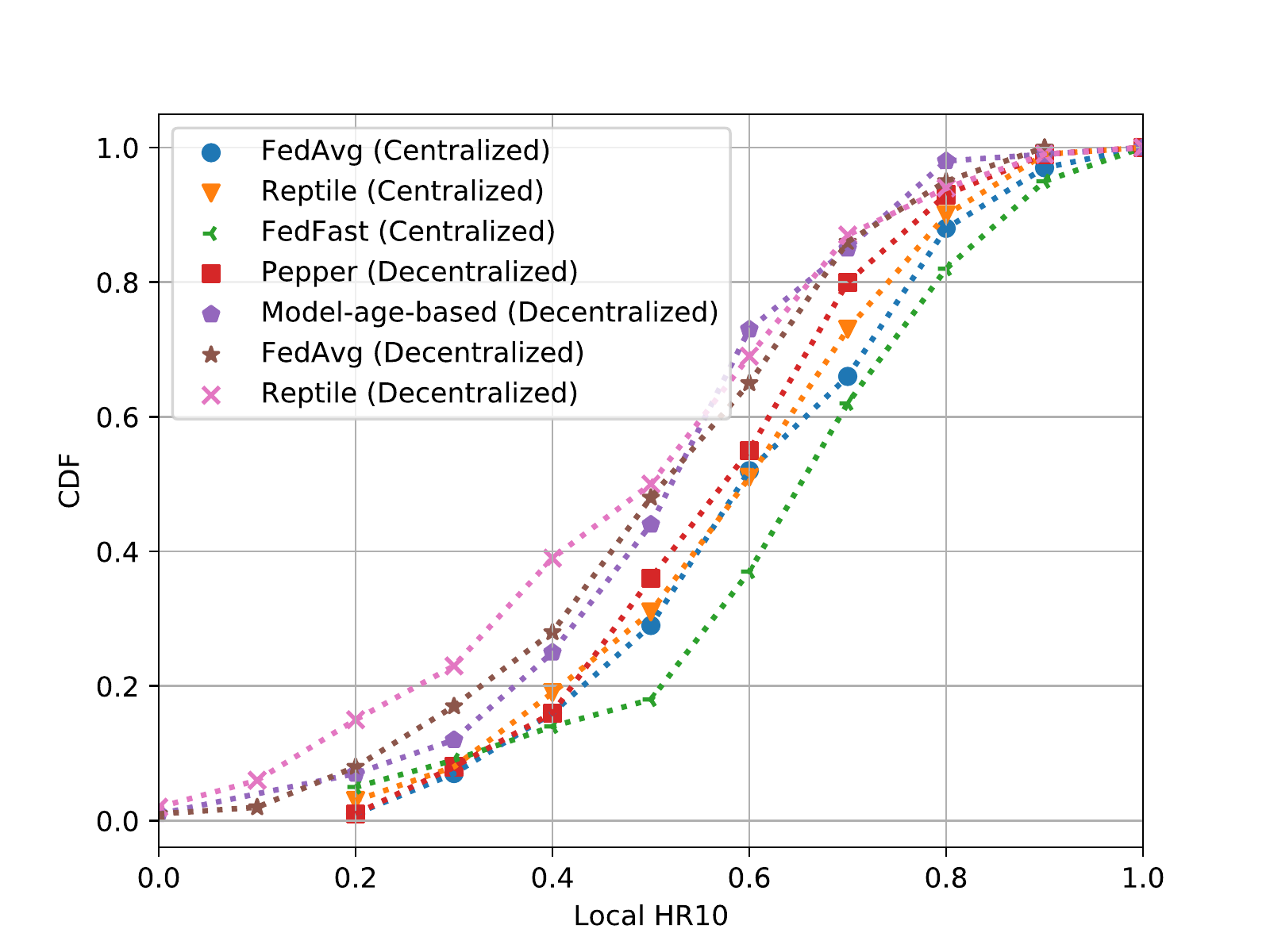}
    \caption{Local Hit Ratio@10 cumulative distribution function.}
    \label{fig:percentilehrmovielens10}
\end{subfigure}
\begin{subfigure}[]{0.45\textwidth}
    \centering
    \includegraphics[width=\textwidth]{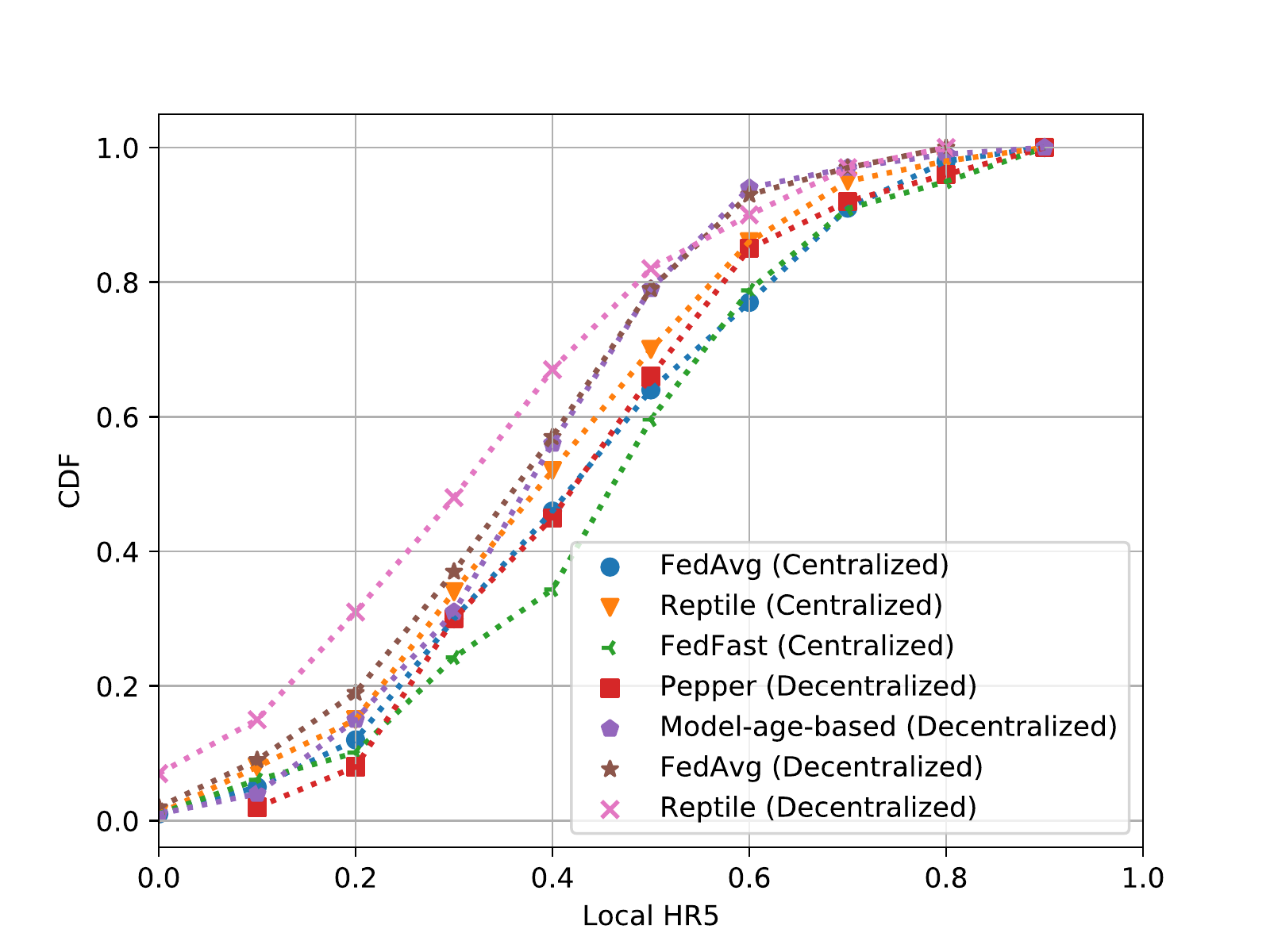}
    \caption{Local Hit Ratio@5 cumulative distribution function.}
    \label{fig:percentilehrmovielens5}
\end{subfigure}
\caption{Top-K recommendation quality distribution comparison on MovieLens\hyp{100K} (GMF, K = 10 and K = 5).}
\end{figure}

\begin{figure}[!htb]
\centering
\begin{subfigure}[]{0.45\textwidth}
    \centering
    \includegraphics[width=\textwidth]{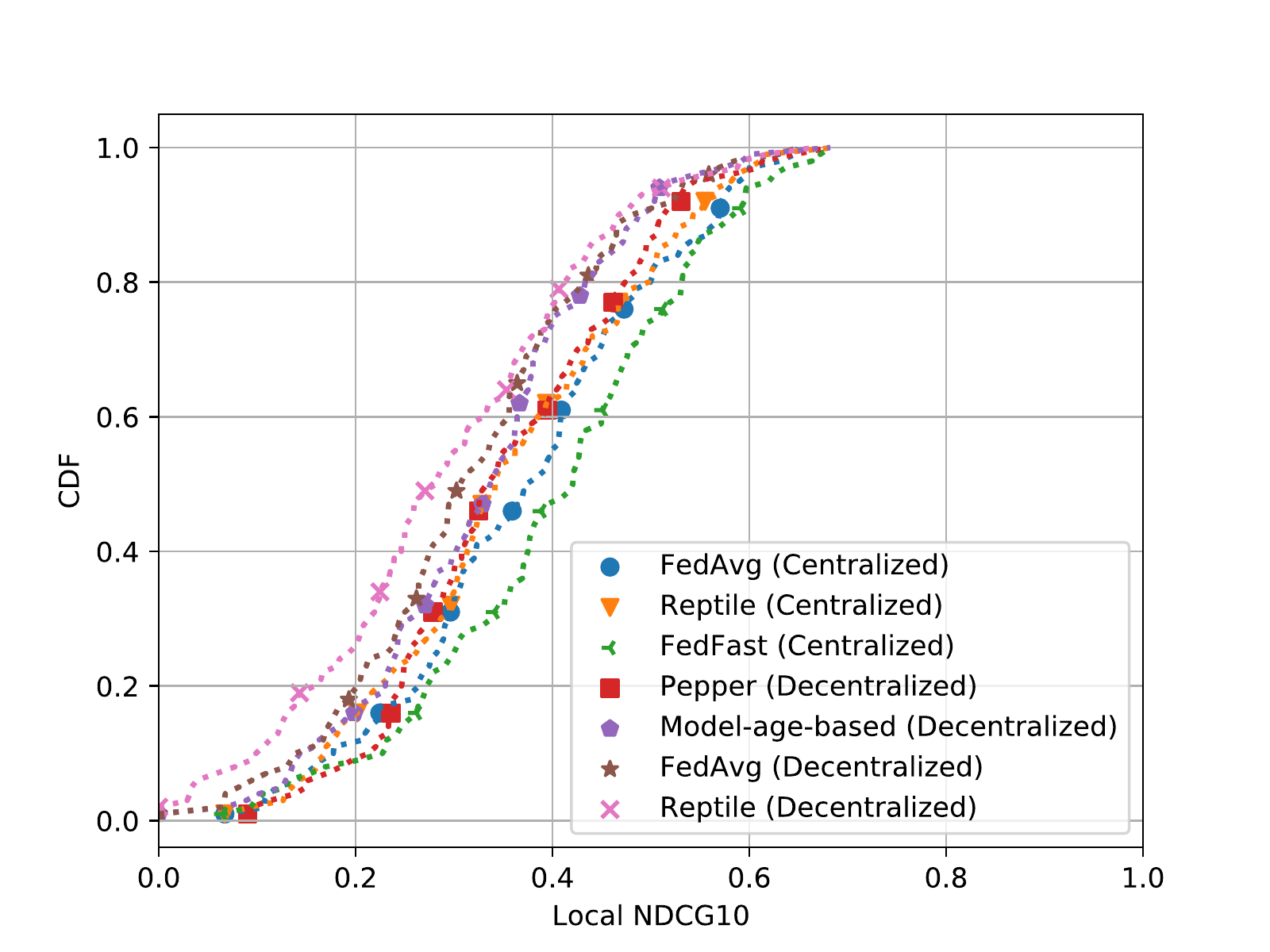}
    \caption{Local NDCG@10 cumulative distribution function.}
    \label{fig:percentilendcgmovielens10}
\end{subfigure}
\begin{subfigure}[]{0.45\textwidth}
    \centering
    \includegraphics[width=\textwidth]{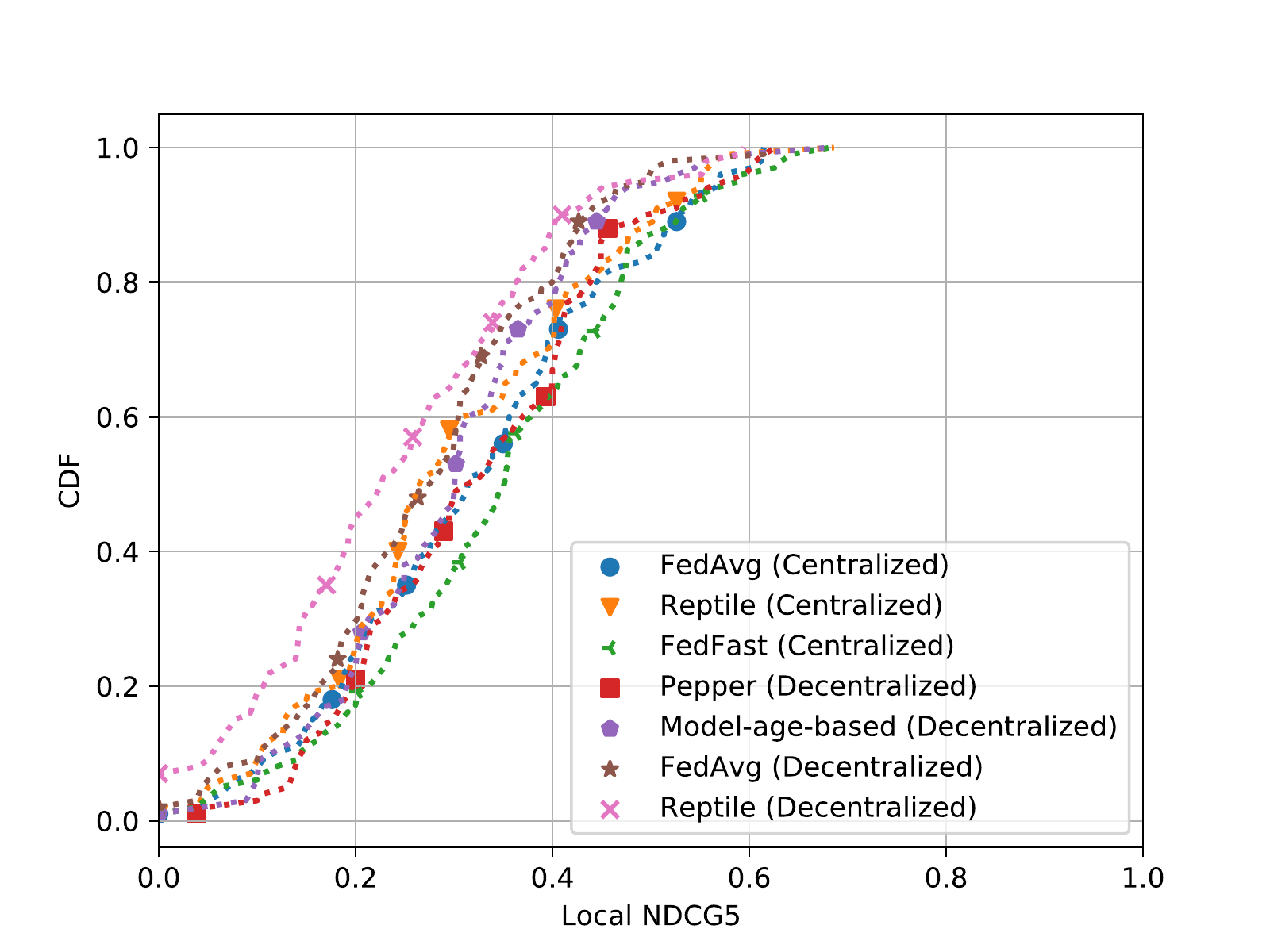}
    \caption{Local NDCG@5 cumulative distribution function.}
    \label{fig:percentilendcgmovielens5}
\end{subfigure}
\caption{ Top-K recommendation quality distribution comparison on MovieLens\hyp{100K} (GMF, K = 10 and K = 5).}
\end{figure}

\begin{figure}[!htb]
\centering
\begin{subfigure}[]{0.45\textwidth}
    \centering
    \includegraphics[width=\textwidth]{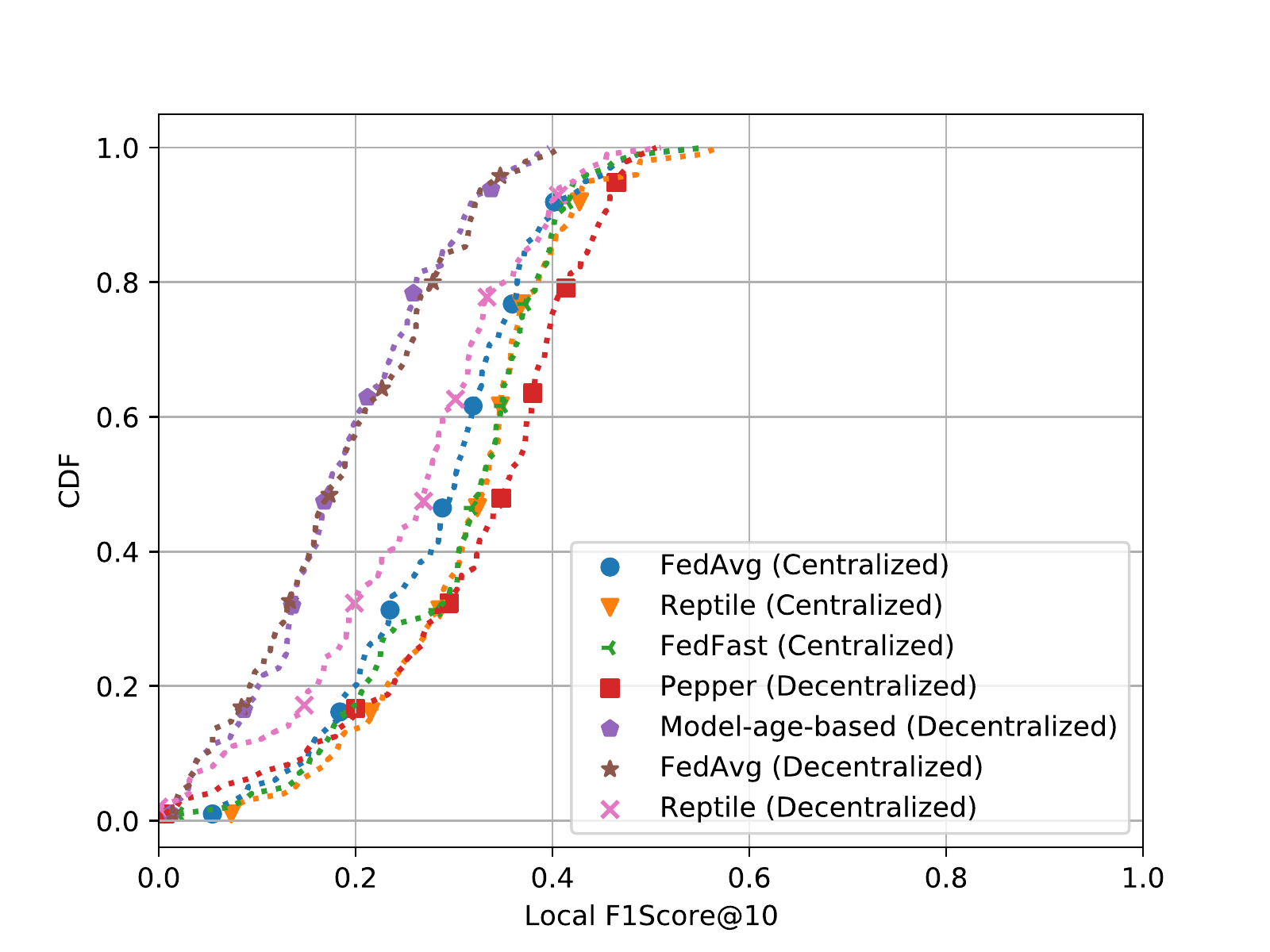}
    \caption{Local F1 score@10 cumulative distribution function.}
    \label{fig:percentilef1scorefoursquare10}
\end{subfigure}
\begin{subfigure}[]{0.45\textwidth}
    \centering
    \includegraphics[width=\textwidth]{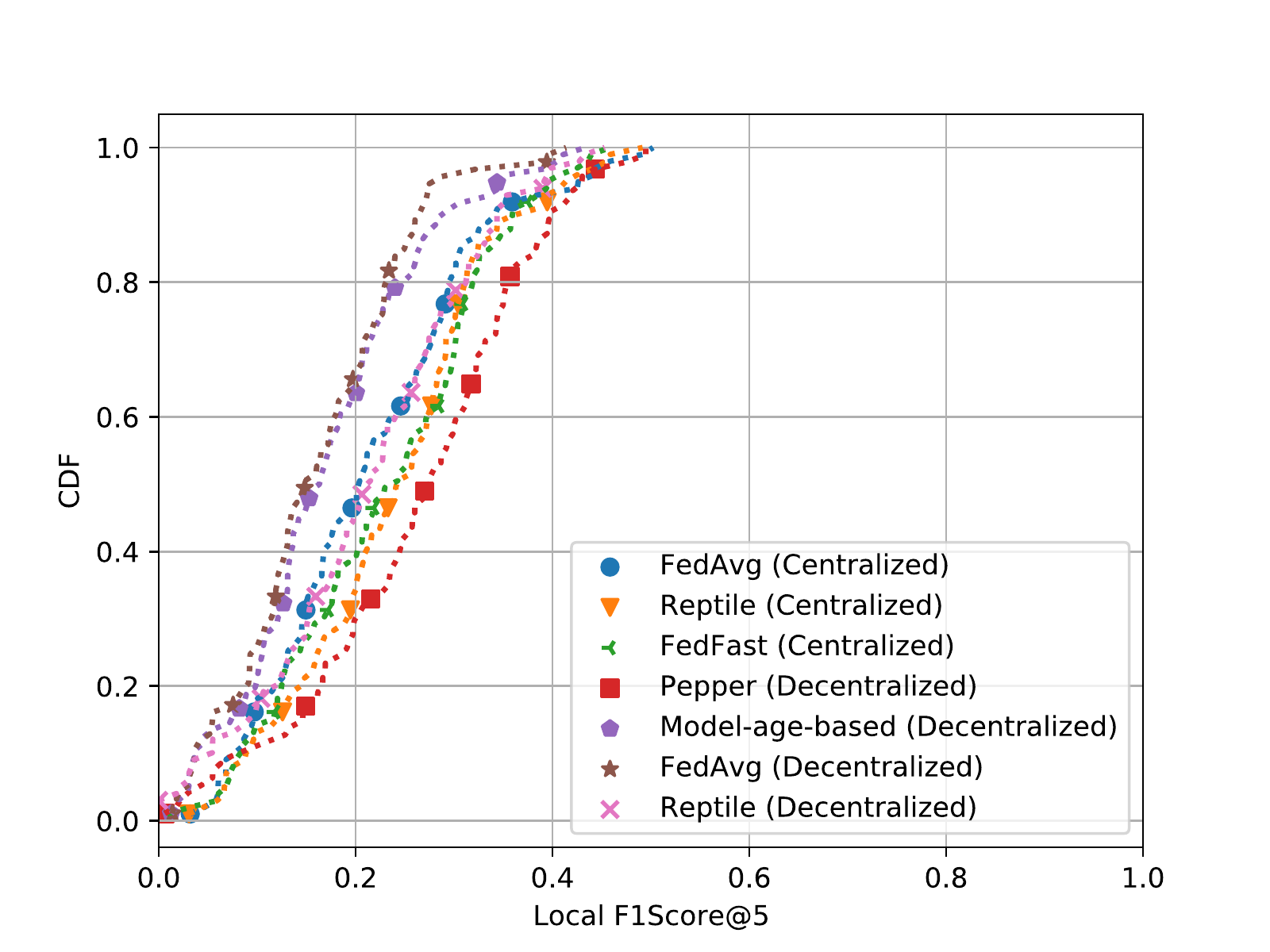}
    \caption{Local F1 score@5 cumulative distribution function.}
    \label{fig:percentilef1scorefoursquare5}
\end{subfigure}
\caption{Top-K F1-Score cumulative distribution function on Foursquare-NYC (PRME-G, K = 10 and K = 5).}
\end{figure}

\begin{figure}[!htp]
\centering
\begin{subfigure}[]{0.45\textwidth}
    \centering
    \includegraphics[width=\textwidth]{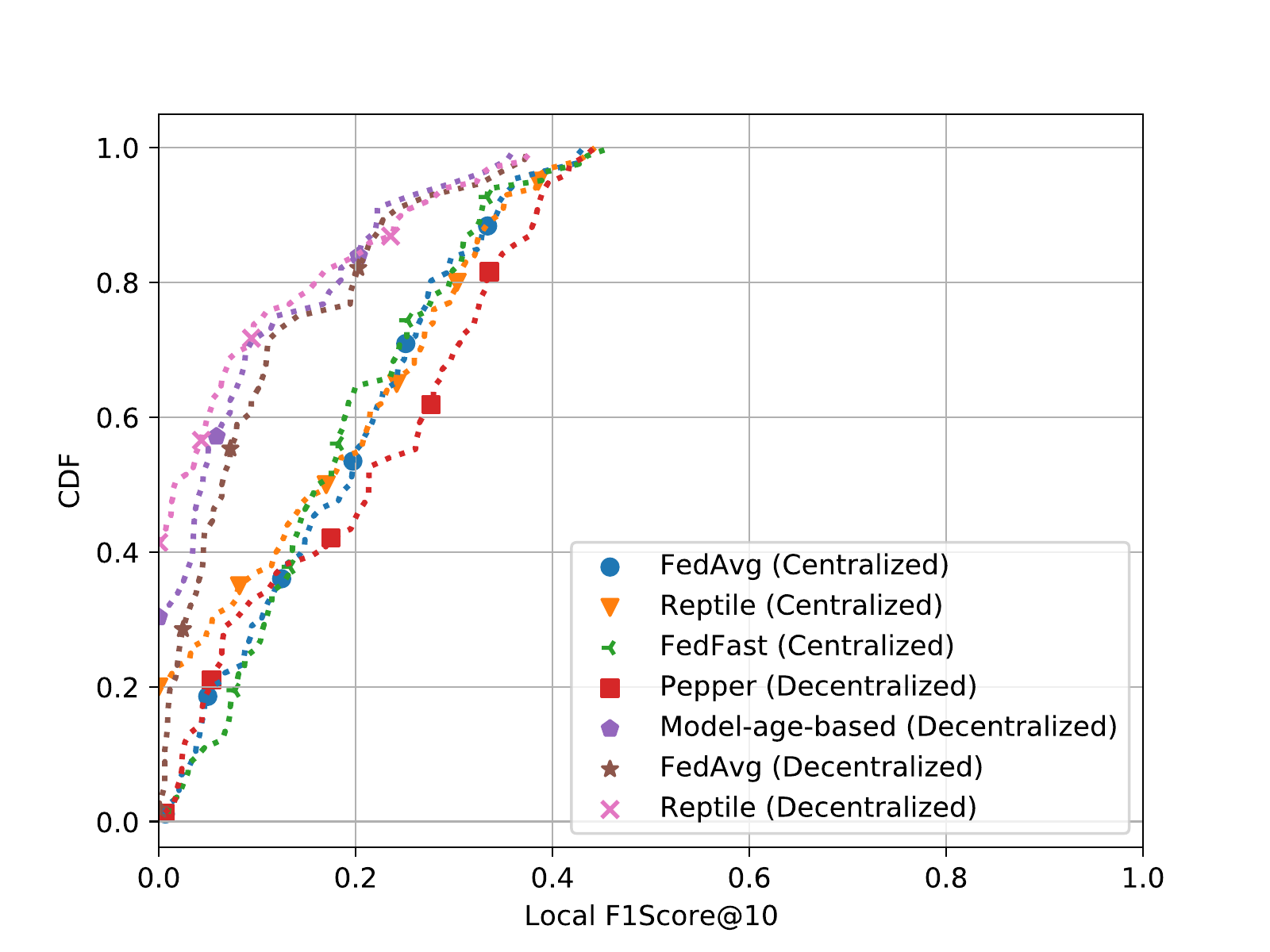}
    \caption{Local F1 score@10 cumulative distribution function.}
    \label{fig:percentilef1scoregowalla10}
\end{subfigure}
\begin{subfigure}[]{0.45\textwidth}
    \centering
    \includegraphics[width=\textwidth]{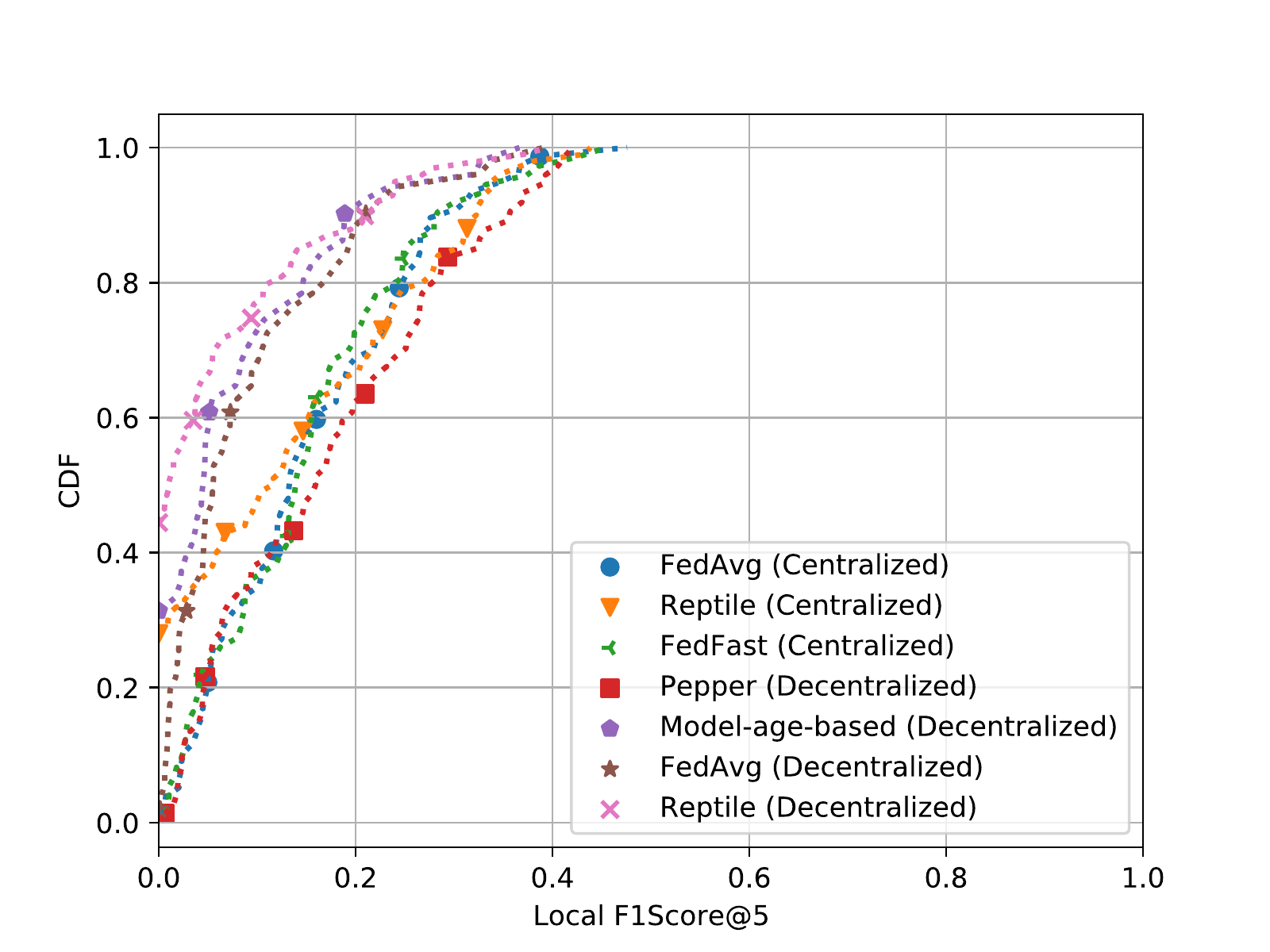}
    \caption{Local F1 score@5 cumulative distribution function.}
    \label{fig:percentilef1scoregowalla5}
\end{subfigure}
\caption{Top-K F1-Score cumulative distribution function on Gowalla-NYC (PRME-G, K = 10 and K = 5).}
\end{figure}


\end{document}